\documentstyle[aps,eqsecnum,epsf,12pt]{revtex}

\newcommand{\Eq}[1]{Eq. (\ref{#1})}
\newcommand{\Eqs}[2]{Eqs. (\ref{#1}) and (\ref{#2})}
\newcommand{\Eqss}[3]{Eqs. (\ref{#1}), (\ref{#2}), and (\ref{#3})}

\newlength{\oldcolwidth}

\begin{document}

\title{\bf Casimir forces in binary liquid mixtures}
\author{Michael Krech}
\address{Fachbereich Physik, Bergische Universit\"at
 Wuppertal, 42097 Wuppertal \\ Federal Republic of Germany}
\maketitle

\begin{abstract}
If two ore more bodies are immersed in a critical fluid critical
fluctuations of the order parameter generate long ranged
forces between these bodies. Due to the underlying mechanism these
forces are close analogues of the well known Casimir forces in
electromagnetism. For the special case of a binary liquid mixture near
its critical demixing transition confined to a simple parallel plate
geometry it is shown that the corresponding critical Casimir forces
can be of the same order of magnitude as the dispersion (van der
Waals) forces between the plates. In wetting experiments or by
direct measurements with an atomic force microscope the resulting
modification of the usual dispersion forces in the critical regime
should therefore be easily detectable. Analytical estimates for the
Casimir amplitudes $\Delta$ in $d=4-\varepsilon$ are compared with
corresponding Monte-Carlo results in $d=3$ and their quantitative
effect on the thickness of critical wetting layers and on force
measurements is discussed.
\end{abstract}

\draft
\pacs{PACS numbers: 64.60.Fr, 05.70.Jk, 68.35.Rh, 68.15.+e}
\setlength{\oldcolwidth}{\columnwidth}

\section{Introduction}
The phase diagram of a fluid is influenced by the presence of a
surface in many different ways. Most prominent is the modification
of the critical behavior of a fluid near a wall \cite{KB83,HWD86} and
the occurrence of new phase transitions induced by the wall such as
wetting and drying \cite{SD88}. For binary liquid mixtures external
walls usually manifest themselves by a preferential affinity of the
wall material for one of the components \cite{MEFHAY80} which in the
vicinity of the critical demixing point leads to the phenomenon of
critical adsorption of the preferred component \cite{HWDAC91,HWDMS93}.
If the system is made finite by the introduction of a second wall or
by confining the system to another finite geometry the critical
behavior of the fluid is modified again if the correlation length
becomes comparable to the system size \cite{MEF70,MEFHN81,MNB83},
where the size dependence of thermodynamic functions takes a scaling
form. A finite geometry may also by generated spontaneously by a
critical fluid if, e.g., a binary liquid mixture near its critical
demixing transition forms a macroscopic wetting layer on the surface
of a substrate \cite{SD88,MPNJOI85}. With the introduction of the
second surface the variety of phenomena in the confined fluid goes far
beyond critical finite-size scaling. Apart from the shift of the
critical point of the system \cite{HNMEF83,KBDPL91} one encounters the
phenomenon of capillary condensation \cite{REUBM86} if the confining
walls of the film geometry consist of the same material. The confinement
of the fluid causes the liquid vapor coexistence line to be shifted
away from the coexistence line of the bulk fluid into the one-phase
regime of, e.g., the bulk vapor \cite{KBDPL91,REUBM86}. For not too
small wall separations a first order phase transition occurs from a
confined vapor to a confined fluid as the undersaturation of the vapor
is lowered at fixed temperature. In a constant-temperature plane of
the phase diagram the line of two-phase coexistence is terminated by a
capillary critical point characterized by a critical undersaturation
and a critical wall separation beyond which capillary condensation no
longer occurs \cite{REUBM86}. Fluid layers growing on the inner walls
of the capillary reduce its effective width and therefore generate
correction terms to the well known Kelvin equation, which describes
the aforementioned shift of the liquid vapor coexistence line as a
function of the width of the capillary \cite{AOPRE92,KBDPL92}.

From the theoretical
point of view these phenomena can be described using density
functional theory \cite{REUBM86} and computer simulations of lattice
gas models \cite{KBDPL91,KBDPL92}. These lattice gases are equivalent
to Ising models, where the presence of the walls is described by {\em
surface fields} which impose a finite surface magnetization on the
Ising system. Density or concentration profiles of confined fluids or
binary liquid mixtures, respectively, then translate to the {\em
magnetization profile} of the Ising model. For the description of
capillary condensation an Ising model with surface fields of the same
sign is appropiate. The behavior of the system changes drastically, if
{\em opposing} surface fields are considered. For a confined binary
liquid mixture this means that the walls perfer {\em different}
components. It turns out that in this case new quasi wetting
transitions occur which can be first-order, critical, and tricritical
and converge to the usual wetting transitions for growing wall
separation \cite{MRSALOJOI91}. Furthermore, two phase coexistence
becomes restricted to temperatures located {\em below} the wetting
temperature, if the surface fields are equal in opposite
\cite{AOPRE92a,KBDPLAMF95}. The scaling behavior of the magnetization
profile of an Ising model with opposing surface fields and the
dependence of the interface position on the strength of the surface
fields and the temperature has been studied thoroughly
\cite{AOP92,KBREDPLAMF96}. Capillary condenstion does no longer occur,
instead one observes the interface delocalization transition, i.e.,
the interface in the magnetization profile detaches from one of the
walls and moves to the midplane of the film. This transition is second
order and its critical point can be identified with the shifted
critical point of the confined system, which in this case is located
on the temperature axis \cite{KBDPLAMF95}. Above the critical
temperature the magnetization profiles become perfectly antisymmetric
about the midplane of the film. By increasing the strength of the
surface fields the critical temperature diminishes and in the limit of
infinitely strong opposing surface fields the interface delocalization
transition becomes suppressed at all.

Confined critical fluids also generate long-ranged forces between the
confining walls \cite{MK94}, a phenomenon, which is a direct analogue of
the well-known Casimir effect in electromagnetism \cite{LS96}. Contrary
to the usual dispersion forces, which are still under investigation for
bodies with curved surfaces \cite{CCS94,GHNIB94,SLAR96} and in presence
of surface roughness \cite{NSC92,BKM95}, critical Casimir
forces are governed by {\em universal scaling functions}
\cite{MNB83,MKSD92}. At the bulk critical point these scaling functions
reduce to the universal Casimir amplitudes \cite{MNB83,MKSD92}. 
Especially for the strip geometry a variety of exact results are known
from conformal invariance \cite{JLC86}. Away from the critical point the
scaling functions are only known exactly for an Ising model confined
to a strip in $d = 2$ \cite{REJS94}. In higher dimensions only the
spherical model has given access to further exact results for the
scaling function of the Casimir force \cite{DD96}. For the $O(N)$
universality class in $d = 3$ so far only approximate results are
known based on real space renormalitation \cite{INW86},
the field theoretic renormalization group \cite{MKSD92}, and
Monte-Carlo simulations \cite{MKDPL96} for the film geometry. More
recently Casimir forces between spherical particles immersed in a
critical $O(N)$ symmetric systems have been investigated by
field-theoretic methods augmented by conformal invariance
considerations \cite{TWBEEUR95}. The field-theoretic treatment of
critical systems confined to finite geometries is notoriously difficult,
because the theory has to interpolate properly between critical behavior
in different dimensions. There has been remarkable progress in devising
alternative renormalization prescriptions beyond the standard minimal
subtraction scheme \cite{DOCCRS94} and in contructing effective actions
for the Ising \cite{Dohm1} and the more general $O(N)$ universality
class \cite{DohmN}. However, these approaches have been devised for
finite systems with symmetry {\em conserving} boundary conditions,
their implementation for systems with symmetry {\em breaking} boundary
conditions (surface fields), in which we are interested here, is still
lacking. Within the framework of Ginzburg-Landau descriptions of
critical finite systems in presence of surface fields the theoretical
treatment has been limited to mean-field considerations for the film
geometry \cite{MEFHN81,HNMEF83,INW86,MIKANO72,BKJZJ93} and concentric
spheres \cite{SGUR95} which can be mapped onto two-sphere and wall-sphere
geometries {\em at} the bulk critical point by conformal transformations
(see Ref. \cite{TWBEEUR95}). For tricritical systems between parallel
plates a thorough mean-field analysis has also been performed
\cite{GG83}. In this paper we will concentrate on the Casimir forces
in critical films in presence of surface fields, which is the adequate
description for confined binary liquid mixtures \cite{MEFHAY80}.

The remainder of the presentation is planned as follows. In Sec. II we
introduce the field-theoretic model of a confined binary liquid mixture
close to its critical demixing point and an adequate Ising model for
which the Monte-Carlo simulations of the Casimir force are performed.
Sec. III is devoted to a survey of mean-field results for the scaling
functions of the Casimir force. In Sec. IV we present one-loop
results and Monte-Carlo estimates for the universal Casimir amplitudes
which characterize the strength of the Casimir forces {\em at} bulk
criticality. We restrict ourselves to the bulk critical point, because
the one-loop calculations are based on the standard
$\varepsilon$-expansion which cannot cope with the dimensional crossover.
In Sec. V we discuss implications of the results presented in Sec. IV
for force measurements and wetting experiments with critical binary
liquid mixtures and we summarize the main results in Sec. VI. The
one-loop calculation requires the knowledge of the mean-field order
parameter profiles which are rederived and discussed in Appendix A.
The eigenmode spectra are derived in Appendix B and the regularization
of the one-loop mode sums is described in Appendix C.

\section{Model}
For the analytical part of the current investigation the standard
$\phi^4$ Ginzburg-Landau Hamiltonian ${\cal H} = {\cal H}_b + {\cal H}_s$
for a $O(N)$ symmetric critical system in a parallel plate geometry
is used. Specifically, the model is defined by the bulk Hamiltonian
\begin{equation}
\label{Hb}
{\cal H}_b[{\bf \Phi}] = \int d^{d-1}x \int_0^L dz \left\{
{1 \over 2} (\nabla {\bf \Phi})^2 + {\tau \over 2} {\bf \Phi}^2 +
{g \over 4!} ({\bf \Phi}^2)^2 \right\},
\end{equation}
where $L$ is the Film thickness, ${\bf \Phi} \equiv (\Phi_1({\bf x},z),
\dots , \Phi_N({\bf x},z))$ is the $N$ component order parameter at the
lateral position ${\bf x}$ and the perpendicular position $z$
$(0 < z < L)$, $\tau$ is the bare reduced temperature, and $g$
is the bare coupling constant. The presence of the surfaces gives rise
to the surface contribution
\begin{equation}
\label{Hs}
{\cal H}_s[{\bf \Phi}] = \int d^{d-1}x \left\{
{c_1 \over 2} [{\bf \Phi}({\bf x},0)]^2 +
{c_2 \over 2} [{\bf \Phi}({\bf x},L)]^2 -
{\bf h}_1 \cdot {\bf \Phi}({\bf x},0) -
{\bf h}_2 \cdot {\bf \Phi}({\bf x},L), \right\}.
\end{equation}
to the Ginzburg-Landau Hamiltonian, where $c_1$ and $c_2$ are
the surface enhancements which characterize the surface
universality class \cite{HWD86}. In mean field theory and within the
dimensional regularization scheme for the field-theoretic
renormalization group $c_i > 0$ defines the {\em ordinary} $(O)$
surface universality class and $c_i < 0$ defines the {\em
extraordinary} $(E)$ surface universality class. The leading critical
behavior of a {\em semiinfinite} system with an $O$ or an $E$ surface is
described by the two {\em stable} renormalization group {\em fixed
point} values $c = +\infty$ and $c = -\infty$, respectively.
Finite positive or negative values of $c_i$ only yield corrections to
the leading behavior. Within this setting $c = 0$ is an {\em unstable}
fixed point, so that $(\tau,c) = (0,0)$ has meaning of a {\em
multicritical} point at which both the bulk and the surface of a
semiinfinite system {\em simultaneously} undergo a second order phase
transition \cite{HWD86}. This mulitcritical point defines a surface
universality class in its own right which is commonly denoted as the
{\em surface-bulk} $(SB)$ or {\em special} universality class. In the
language of a spin model $c$ denotes the deviation of the exchange
interaction between spins in the surface from its value at the
multicritical point [see also \Eq{HIsing} below].

The quantities ${\bf h}_1$ and ${\bf h}_2$ denote surface fields which
explicitly break the $O(N)$ symmetry of the model. In case of a broken
symmetry at the surface in principle also cubic surface fields need to
be considered \cite{HWDAC91}. However, for the investigation of the
{\em leading} critical behavior in the presence of nonzero linear
surface fields cubic surface fields can be disregarded \cite{HWDAC91}.

As pointed out in Sec. I a wall which is in contact with a binary
liquid mixture will in general show some preferential affinity for one
of the components so that the composition profile varies as a function
of the perpendicular coordinate $z$. This situation can be represented
by setting $c_1 \geq 0$ and $c_2 \geq 0$ in \Eq{Hs} and prescribing
finite values for the surface fields ${\bf h}_1$ and ${\bf h}_2$. The
phase transition in the bulk in presence of nonzero surface fields is
called the {\em normal} transition \cite{TWBHWD94}. As far as the
leading critical behavior is concerned the normal transition is
equivalent to the usual extraordinary transition \cite{HWD86,TWBHWD94},
which can be represented by setting ${\bf h}_1 = {\bf h}_2 = 0$ and
choosing $c_1 < 0$ and $c_2 < 0$. In the following we will therefore
exclusively use the surface field picture of the extraordinary
transition.

In the field-theoretic analysis only the cases of strictly parallel and
strictly antiparallel surface fields ${\bf h}_i = (h_i,0,\dots ,0)$,
$i=1,2$ will be considered. For the leading critical behavior it is
sufficient to discuss only the limiting cases $h_1, h_2 \to \pm \infty$
\cite{HWD86}. The above restriction to parallel and antiparallel
surface fields then means that we only consider the two cases $h_1 =
h_2 \to +\infty$ and $h_1 = -h_2 \to +\infty$. To simplify the
notation we will refer to the former case as the $(+,+)$ boundary
condition and to the latter case as the $(+,-)$ boundary condition
which are the only combinations $(E,E)$ of the $E$ surface
universality class in the film geometry considered here. One can also
combine a symmetry breaking $E$ surface with a symmetry conserving
$O$ or $SB$ surface. However, as will be demonstrated below, the
combinations $(O,E)$ and $(SB,E)$ can be extracted from the analysis of
the cases $(+,-)$ and $(+,+)$, respectively.

For the numerical part of this investigation we restrict ourselves to
the case $N=1$ which is the most interesting one in view of
applications of the results to binary liquid mixtures. The simulations
are performed for a spin - $1 \over 2$ Ising model confined to a film
geometry in $d=3$ dimensions defined by the Hamiltonian
\begin{equation}
\label{HIsing}
{\cal H}_I = -J \sum_{<({\bf x},z),({\bf x'},z')>}
s({\bf x},z) s({\bf x'},z') - H_1 \sum_{\bf x} s({\bf x},1)
- H_2 \sum_{\bf x} s({\bf x},L) ,
\end{equation}
where $J$ is the excange coupling constant, $<({\bf x},z),({\bf x'},z')>$
denotes a nearest neighbor pair of spins and the spins $s({\bf x},z)$ can
take the values 1 and $-1$. The underlying lattice is supposed to be
simple cubic with $L'$ lattices sites and periodic boundary conditions
in the $x$ and $y$ directions. In the $z$ direction the lattice has
$L \ll L'$ sites and the missing bonds in the two surface layers at
$z=1$ and $z=L$ are left open. In order to simulate the model at the
normal transition \Eq{HIsing} contains two surface terms by which the
spins in the two surface layers are coupled to surface fields $H_1$
and $H_2$, respectively. Infinite surface fields are simply realized by
fixing all spins in the surface to a fixed value 1 or $-1$ depending on
the sign of the surface field. In the model defined by \Eq{HIsing} the
surface exchange coupling constant $J_1$ has the fixed value $J_1 = J$.
It has been shown by Monte-Carlo simulations of spin - $1 \over 2$ Ising
models that the $SB$ multicritical point is characterized by the
special value $J_{1c} \simeq 1.50 J$ \cite{CRFW95} of the surface
coupling constant $J_1$. Apart from corrections to scaling the $O$
surface universality class is represented by the condition $J_1 < J_{1c}$
\cite{HWD86} which is fulfilled by \Eq{HIsing} due to $J_1 = J < J_{1c}$.
Therefore only the $O$ surface universality class ($H_1 = 0$ or
$H_2 = 0$) and the $E$ surface universality class ($H_1 \neq 0$ or
$H_2 \neq 0$) can be studied with the above Ising model Hamiltonian.
The film geometry underlying \Eq{HIsing} then allows the investigation
of the four combinations $(O,O)$, $(O,E)$, $(+,+)$, and $(+,-)$ of
boundary conditions by a Monte-Carlo simulation, where the
combination $(O,E)$ means $(O,+)$ or, equivalently, $(O,-)$. The
principal setup of a Monte-Carlo algorithm for a measurement of the
Casimir force in lattice models is described in Ref. \cite{MKDPL96}
to which the reader is referred for further details.

\section{Landau theory}
The presence of a symmetry breaking surface field implies a
nonvanishing order parameter profile for all $\tau$ (see Appendix A),
which substantially complicates the field theoretic analysis of the
Casimir effect as compared to the case of symmetry conserving boundary
conditions discussed in Ref. \cite{MKSD92}. On the other hand the
leading (mean field) contribution to the Casimir amplitude can be
determined without any detailed knowledge about the functional form of
the order parameter profile. We briefly illustrate this for the case
$\tau = 0$ and $N=1$, i.e., ${\bf \Phi} = (\Phi,0,\dots,0)$ in
\Eqs{Hb}{Hs}. In the mean field approximation the order parameter
profile has the form $\overline{\bf \Phi}({\bf x},z) = (M(z),0,\dots,0)$,
where $M(z)$ solves the Euler-Lagrange equations given by \Eqs{EL}{ELbc}.
Inserting $\overline{\bf \Phi}$ into \Eqs{Hb}{Hs} for $\tau = 0$ and
integrating by parts using \Eqss{EL}{ELbc}{ELint} ${\cal H}[\overline{\bf
\Phi}]$ can be evaluated without solving the Euler-Lagrange equations
for $M(z)$ explicitly. The result is the mean field free energy of the
film at bulk criticality and is given by
\begin{eqnarray}
\label{HPhi}
{\cal H}[\overline{\bf \Phi}] &=&
{c_1 \over 6} M^2(0) + {c_2 \over 6} M^2(L) -
{2 \over 3} h_1 M(0) - {2 \over 3} h_2 M(L)
\nonumber \\ \\
&+& {L \over 3} \left[{1 \over 2} M'^2(z_0) - {g \over 4!} M^4(z_0)
\right] \nonumber ,
\end{eqnarray}
where $h_1$ and $h_2$ denote the first component of ${\bf h}_1$ and
${\bf h}_2$, respectively, and $z_0$ is an arbitrary reference point
$0 \leq z_0 \leq L$ between the two surfaces of the film. The terms
in the first line of \Eq{HPhi} constitute the surface contribution to
the mean field free energy and the contribution in the second line of
\Eq{HPhi} is the finite size part, where the square bracket yields the
Casimir force (see below). As a direct implication of \Eq{ELint} one
finds that the above expression for the Casimir force does not depend
on the reference point $z_0$.
Note that due to $\tau = 0$ the bulk contribution to \Eq{HPhi} vanishes
identically. For $\tau \neq 0$ ${\cal H}[\overline{\bf \Phi}]$ cannot
be expressed in the same closed form as given by \Eq{HPhi} and we
therefore resort to the $zz$-component $T_{\perp \perp}$ of the {\em
stress tensor} $T_{kl}({\bf x},z)$ in order to find a
more general expression for the Casimir force. The stress tensor
$T_{kl}$ is given by \cite{EEMS94}
\begin{eqnarray}
\label{Tkl}
T_{kl} &=& {\partial {\bf \Phi} \over \partial x_k}
\cdot {\partial {\bf \Phi} \over \partial x_l} - \delta_{kl} \left[
{1 \over 2} (\nabla {\bf \Phi})^2 + {\tau \over 2} {\bf \Phi}^2 +
{g \over 4!} ({\bf \Phi}^2)^2 \right] \nonumber \\ \\
&-& \left[{d-2 \over 4(d-1)} + {\cal O}(g^3) \right] \left[
{\partial^2 \over \partial x_k \partial x_l} - \delta_{kl} \nabla^2
\right] {\bf \Phi}^2 , \nonumber
\end{eqnarray}
where $\tau$ and $g$ have the same meaning as in \Eq{Hb}.
The scaling dimension of $T_{kl}$ is given by the spatial
dimension $d$. In a film geometry $\langle T_{kl} \rangle$ is diagonal
due to the lateral translational invariance of the film. From the
conservation property $\partial \langle T_{kl} \rangle / \partial x_k
= 0$ one then concludes that $\langle T_{kl} \rangle$ does not depend
on position and therefore $\langle T_{\perp \perp} \rangle$
can be directly identified with the Casimir force per
unit area. Note that the evaluation of $\langle T_{kl} \rangle$
according to \Eq{Tkl} for $x_k = x_l = z$ and for $\tau = 0$ within
the mean field approximation $\overline{\bf \Phi}({\bf x},z) =
(M(z),0,\dots ,0)$ for the order parameter yields the square bracket
in \Eq{HPhi}.

We now turn to the mean field analysis of the Casimir force as a
function of the reduced temperature $\tau$, where we first restrict
ourselves to the case $N = 1$ (Ising universality class). In view of
later applications of the results to binary liquid mixtures near the
critical demixing transition this is the most relevant case. For the
mean field analysis alone it would not be neccessary to determine the
full order parameter profiles. However, in order to perform the
fluctuation expansion (see Sec. IV and Appendix B) precise knowledge
about the profiles on the mean field level is indispensable. Details
of the calculation are summarized in Appendix A. In the course of the
calculations for the order parameter profiles one obtains the
corresponding expressions for the Casimir forces as byproducts which
will be discussed in the following paragraph.

As in Appendix A we write the mean field contribution $\langle
T_{\perp \perp} \rangle_0$ to the Casimir force in the form $\langle
T_{\perp \perp} \rangle_0 = (6/g) t_{\perp \perp}$ and we only
consider $t_{\perp \perp}$ in the following for simplicity. From the
general theory of critical finite size scaling \cite{MEF70,MNB83} we
expect $t_{\perp \perp}$ to take the scaling form
\begin{equation}
\label{tscal}
t_{\perp \perp} = L^{-d} F(y)\ , \ y = \tau L^{1/\nu} ,
\end{equation}
where $d = 4$ and $\nu = 1/2$ within mean field theory. Note that
right {\em at} the upper critical dimension $d = d_c = 4$ the
prefactor $6/g$ of the Casimir force generates logarithmic finite-size
corrections due to the fact that the renormalized
counterpart $u$ of the coupling constant $g$ vanishes according to
$u(l) \sim 1/\ln l$ for $l \to \infty$ at the renormalization group
fixed point \cite{JRHGDJ85}. However, logarithmic corrections to
scaling in $d = 4$ will be disregarded here so that from the point of
view of mean field theory the above prefactor is treated as a constant.

\begin{figure}[t]
\centerline{\epsfbox{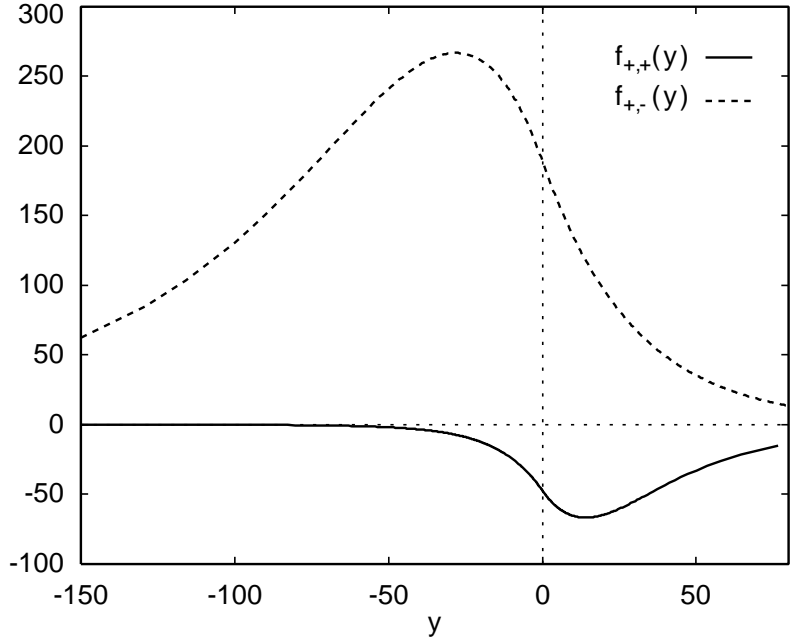}}
\centerline{\caption{\protect\small{
Scaling functions $f_{+,+}(y)$ (solid line) and $f_{+,-}(y)$ (dashed
line) according to \protect\Eqss{fpp}{fpm}{fab}. The $y$ range
influenced by the bulk critical point $y = 0$ is very broad and the
asymptotic decay for $y \to \pm \infty$ is dominated by an exponential
[see \protect\Eq{fabasy}]. Note that $f_{+,+}(y)$ and $f_{+,-}(y)$ take
their extreme values at $y \simeq 10$ and $y \simeq -25$, respectively.
\label{fabplot}}}}
\end{figure}

For the case of $(+,+)$ boundary conditions the scaling function
$F_{+,+}(y)$ can be read off from \Eqs{tautpp}{tautpp1}. The result is
\begin{eqnarray}
\label{fpp}
F_{+,+}(y) = -(2K)^4 k^2 (1 - k^2) &\quad , \quad&
y = (2K)^2 (2k^2 - 1) ; \nonumber \\
F_{+,+}(y) = (2K)^4 k^2 &\quad , \quad&
y = -(2K)^2 (k^2 + 1) ,
\end{eqnarray}
where $K \equiv K(k)$ is the complete elliptic integral of the first
kind and $0 \leq k < 1$. The $y$ dependence of $F_{+,+}$ according to
\Eq{fpp} is given in the parametric form $y = y(k)$, where $y(k)$ is a
monotonic function of $k$ so that the inverse $k = k(y)$ exists and
constitutes the $y$ dependence of $F_{+,+}$ in a unique way. As can be
seen from \Eq{fpp} the parameterizations of $F_{+,+}$ and $y$ for $y \geq
-\pi^2$ and $y \leq -\pi^2$ are different. The reason for this
is purely technical in the sense that negative values for $k^2$ are
avoided (see Appendix A for details). There is no singularity of
$F_{+,+}(y)$ at the point $y = -\pi^2$ ($k = 0$). In fact,
$F_{+,+}(y)$ is analytic for {\em all} values of $y$, because the
critical point of the film $(T_c(L),h_c(L))$ is located {\em off} the
temperature axis at a finite critical bulk field $h = h_c(L) \sim
L^{-\Delta/\nu}$, where $\Delta$ is the gap exponent
\cite{MEFHN81,HNMEF83}. Within mean field theory one has $\Delta =
3/2$ so that $\Delta/\nu = 3$. For $(+,-)$ boundary conditions the
corresponding result for the scaling function $F_{+,-}(y)$ can be
read off from \Eqs{tautpm}{tautpm1}. One finds
\begin{eqnarray}
\label{fpm}
F_{+,-}(y) = (2K)^4 (1 - k^2)^2 &\quad , \quad&
y = 2 (2K)^2 (k^2 + 1) ; \nonumber \\
F_{+,-}(y) = (2K)^4 &\quad , \quad&
y = -2 (2K)^2 (2k^2 - 1) ,
\end{eqnarray}
where a parameterization analogous to the one in \Eq{fpp} has been
used. The scaling function $F_{+,-}(y)$ is also analytic for all
values of y, although the critical point of the film in the case
of opposing surface fields is located on the temperature axis and is
associated with the interface delocalization transition
\cite{KBDPLAMF95}. However, due to the limit $h_1 = -h_2 \to \infty$
performed here this critical point has been formally shifted to $y_c
= -\infty$ so that it is no longer visible as a singularity in
$F_{+,-}(y)$. Corresponding results for $(SB,+)$ and $(O,+)$ boundary
conditions can be constructed from \Eqs{fpp}{fpm} using the simple
transformation $L \to 2L$ (see Appendix A). One obtains
\begin{eqnarray}
\label{fmix}
F_{SB,+}(y) = \textstyle{1 \over 16} F_{+,+}(4y) &\quad , \quad&
F_{O,+}(y) = \textstyle{1 \over 16} F_{+,-}(4y) .
\end{eqnarray}
The scaling functions obtained so far still contain a bulk contribution
which corresponds to a bulk pressure given by $t_{\perp \perp, bulk} =
-\tau m_b^2 - m_b^4$. For $\tau \geq 0$ one has $t_{\perp \perp, bulk} =
0$ $(m_b = 0)$ and for $\tau < 0$ one has $t_{\perp \perp, bulk} =
\tau^2/4$ $(m_b = \sqrt{-\tau/2})$. The bulk contribution
$F_{bulk}(y)$ to the scaling functions given by \Eqss{fpp}{fpm}{fmix} then
has the simple form $F_{bulk}(y) = \theta(-y)\ y^2/4$ which contains the
usual mean field singularity of the bulk free energy at $\tau = 0$.
In order to express the finite-size contribution to the Casimir force
$\langle T_{\perp \perp} \rangle_0$ in the scaling form we define the
scaling functions
\begin{equation}
\label{fab}
f_{a,b}(y) \equiv F_{a,b}(y) - F_{bulk}(y)
\end{equation}
which are displayed in Fig. \ref{fabplot} for $(a,b) = (+,+)$ and
$(+,-)$. Their shapes resemble those of the corresponding scaling
functions for the Ising model confined to a strip in $d = 2$
\cite{REJS94}. The asymptotic behavior of the scaling functions for $y
\to \pm \infty$ is governed by an exponential decay according to
\begin{eqnarray}
\label{fabasy}
f_{+,+}(y \to \infty) &\simeq& -16 y^2 \exp(-\sqrt{y}) , \nonumber \\
f_{+,+}(y \to -\infty) &\simeq& -16 y^2 \exp(-\sqrt{-2y}) , \nonumber \\
f_{+,-}(y \to \infty) &\simeq& 16 y^2 \exp(-\sqrt{y}) , \nonumber \\
f_{+,-}(y \to -\infty) &\simeq& 16 y^2 \exp(-\sqrt{-y/2}) .
\end{eqnarray}
The scaling functions take quite sizable values over a surprisingly
broad range of the scaling argument $y$. This may serve as a first
indication that the Casimir forces provide a strong modification of
the usual dispersion forces in a parallel plate geometry at the
extraordinary transition. However, in order to estimate the absolute
strength of the Casimir forces in binary liquid mixtures close to
their critical demixing transition a renormalization group analysis of
$f_{+,+}(y)$ and $f_{+,-}(y)$ is required (see Sec. IV).
\setlength{\columnwidth}{14.0cm}
\setlength{\columnwidth}{\oldcolwidth}

\begin{figure}[t]
\centerline{\epsfbox{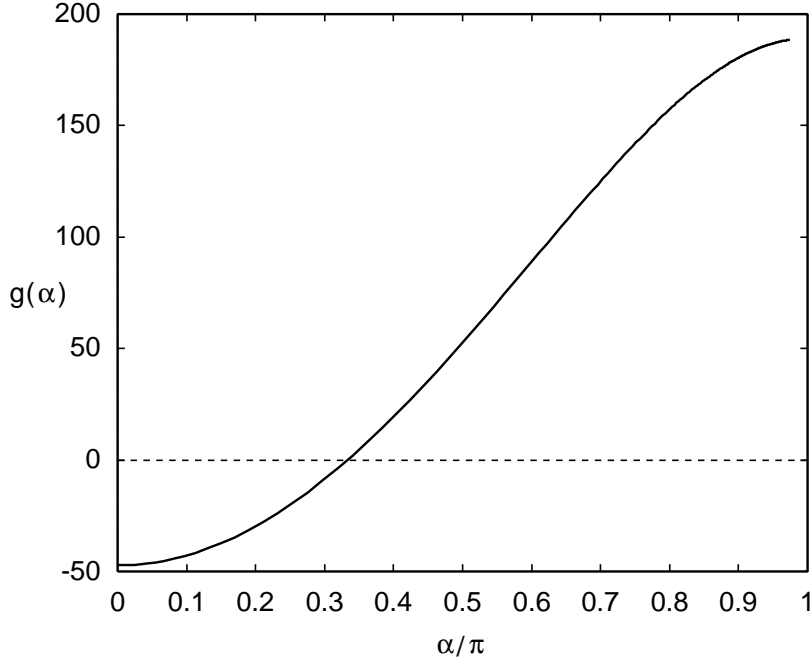}}
\centerline{\caption{\protect\small{
Amplitude function $g(\alpha)$ of the Casimir force according to
\protect\Eqs{talp}{ga}. $g(\alpha)$ smoothly interpolates between
$g(0) = f_{+,+}(0)$ and $g(\pi) = f_{+,-}(0)$ (see main text). The
amplitude vanishes at $\alpha = \pi /3$.
\label{galp}}}}
\end{figure}

If the order parameter has $N > 1$ components, the case of parallel
surface fields is already covered by the above analysis of the $(+,+)$
boundary conditions for $N = 1$, because in this case the order
parameter only has one nonzero component parallel to the surface
fields. For antiparallel surface fields, however, this is not as
obvious, because the order parameter has the additional freedom to
rotate across the film by a position dependent angle $\varphi(z)$. 
We illustrate this for the case $N \geq 2$ with ${\bf h}_1 =
(h_1,0,\dots ,0)$ and ${\bf h}_2 = h_1\ (\cos \alpha, \sin
\alpha,0,\dots ,0)$ in the limit $h_1 \to \infty$ and for $\tau = 0$.
A similar situation has been been discussed in Ref. \cite{BKJZJ93} for
$\tau \neq 0$. If the order parameter profile is written in the form
${\bf M}(z) = \sqrt{12/g}\ m(z)\ (\cos \varphi(z), \sin
\varphi(z),0,\dots ,0)$ one finds the Euler-Lagrange equations [see
\Eq{ELm} and Ref. \cite{BKJZJ93}]
\begin{eqnarray}
\label{ELmphi}
\left[\varphi'(z) m^2(z)\right] ' &=& 0 , \nonumber \\
m''(z) &=& \varphi'^2(z) m(z) + 2m^3(z) .
\end{eqnarray}
As boundary conditions for $\varphi(z)$ we choose $\varphi(0) = 0$ and
$\varphi(L) = \alpha$, because the order parameter should be parallel
to ${\bf h}_1$ and ${\bf h}_2$, respectively, at the surfaces. The
amplitude function $m(z)$ is positive and its qualitative behavior
resembles that of the profile $m_{+,+}(z)$ [see \Eqs{mpp}{mpp1}].
From \Eq{Tkl} we then find for the Casimir force
\begin{equation}
\label{T0phi}
\langle T_{\perp \perp} \rangle_0 \equiv (6/g)\ t_{\perp \perp} =
(6/g)\ \left[\varphi'^2(L/2) m^2(L/2) - m^4(L/2)\right] .
\end{equation}
Note that \Eq{T0phi} allows a sign change of the Casimir force as
a function of the angle $\alpha$ enclosed by the surface fields.
Following Appendix A [see \Eqs{ELint}{tpp}] the first integral of
\Eq{ELmphi} is given by
\begin{eqnarray}
\label{ELmphint}
\varphi'(z) &=& c / m^2(z) , \nonumber \\
m'^2(z) &=& -c^2 / m^2(z) + m^4(z) + t_{\perp \perp},
\end{eqnarray}
where
\begin{equation}
\label{tppc}
t_{\perp \perp} = c^2 / m^2(L/2) - m^4(L/2)
\end{equation}
and $c$ is a constant such that
\begin{equation}
\label{alpc}
\alpha = \varphi(L) \quad \mbox{for} \quad
\varphi(z) = c \int_0^z dz' / m^2(z') .
\end{equation}
Just as for \Eq{P} we apply the substitution $P(z) \equiv m^2(z)$ and
eliminate $c$ using \Eq{tppc}. All the information needed to calculate
the Casimir force, i.e., $t_{\perp \perp}$ as a function of $\alpha$
is now contained in \Eq{alpc} and
\begin{equation}
\label{Palp}
P'^2(z) = 4 \left[ P^3(z) - P^3(L/2) + t_{\perp \perp}
\left(P(z) - P(L/2)\right) \right]
\end{equation}
which shows that $P(z) \equiv \wp(z;g_2,g_3)$ is a Weierstrass
elliptic function, where the invariants $g_2$ and $g_3$ can be read
off from \Eq{Palp}. As we are focussing on the limit of infinite
surface fields the film thickness $L$ is one of the basic periods of
$P(z)$ [see \Eqs{om1om2}{om1om2pm}], and therefore $P(z)$ has double
poles at $z = 0$ and $z = L$. Using \Eq{Palp} we can rewrite \Eq{alpc}
and find a representation for $P(L/2)$ by performing a separation of
variables in \Eq{Palp}. Writing $t_{\perp \perp}$ in the scaling form
[see \Eq{tscal}]
\begin{equation}
\label{talp}
t_{\perp \perp} = L^{-4} g(\alpha)
\end{equation}
and using the abbreviation $p \equiv L\sqrt{P(L/2)}$ one finds
\begin{eqnarray}
\label{ga}
p &=& \int_1^\infty \left[x^3 - 1 + (x - 1) p^{-4}
g(\alpha) \right]^{-1/2} dx , \nonumber \\ \\
\alpha &=& \sqrt{1 + p^{-4} g(\alpha)} \int_1^\infty x^{-1} \left[
x^3 - 1 + (x - 1) p^{-4} g(\alpha) \right]^{-1/2} dx \nonumber .
\end{eqnarray}
The solution of \Eq{ga} is shown in Fig. \ref{galp}.
The Casimir force (i.e., $g(\alpha)$) grows monotonically from $\alpha =
0$ to $\alpha = \pi$ at fixed $L$ and vanishes for the angle $\alpha =
\pi / 3$ which can also be derived directly from \Eq{ga} by setting
$g(\alpha) = 0$. Furthermore it should be noted that according to
\Eqs{fpp}{fab} one has $g(0) = F_{+,+}(0) = f_{+,+}(0)$ and according
to \Eqs{fpm}{fab} one also has $g(\pi) = F_{+,-}(0) = f_{+,-}(0)$. The
function $g(\alpha)$ therefore smoothly interpolates between $(+,+)$
and $(+,-)$ boundary conditions giving the same result for the
Casimir force as the Ising universality class $(N = 1)$ in these two
cases.
\setlength{\columnwidth}{14.0cm}
\setlength{\columnwidth}{\oldcolwidth}

We close this section with a short discussion of the analytic solution
of \Eqs{Palp}{ga}. Following the derivation described in Appendix A
and using \Eq{alpc} the profile $m(z)=\sqrt{P(z)}$, the amplitude
function $g(\alpha)$, and the angle $\alpha$ can be parameterized in
terms of the modulus $k$ of the Jacobian elliptic functions. One finds
\begin{eqnarray}
\label{P1g1}
m(z) &=& {2K \over L} \left[ {\mbox{dn}^2(\zeta;k)
\over \mbox{sn}^2(\zeta;k)} + {2k^2 - 1 \over 3} \right]^{1/2} ,
\nonumber \\ \nonumber \\
g(\alpha) &=& -{1 \over 4} (2K)^4 \left[ 1 + {(2k^2 - 1)^2 \over 3}
\right] , \\ \nonumber \\
\alpha &=& 2 \left[{(1-2k^2)(2-k^2) \over 3 (1+k^2)}\right]^{1/2}
\left[ \Pi(1/3 + k^2/3,k) - K \right] \nonumber
\end{eqnarray}
and
\begin{eqnarray}
\label{P2g2}
m(z) &=& {2K \over L} \left[ {\mbox{cn}^2(\zeta;k)
\over \mbox{sn}^2(\zeta;k)\mbox{dn}^2(\zeta;k)}
- {2(2k^2 - 1) \over 3} \right]^{1/2} , \nonumber \\ \nonumber \\
g(\alpha) &=& (2K)^4 \left[ 1 - {4 (2k^2 - 1)^2 \over 3}
\right] , \nonumber \\ \\
\alpha &=& \sqrt{6(1-2k^2)}
\left\{ \left[ {b \over 1-2k^2} + {2 \over 3} \right]
\left[ \Pi(a,k) - K \right] \right. \nonumber \\ \nonumber \\
&-& \left. \left[ {a \over 1-2k^2} + {2 \over 3} \right]
\left[ \Pi(b,k) - K \right] \right\} , \nonumber
\end{eqnarray}
where $\zeta = (2K/L)z$, $0 \leq k^2 \leq 1/2$, and the parameters $a$
and $b$ are given by
\begin{eqnarray}
\label{ab}
a &=& \textstyle{1 \over 6} \left[1 + 4k^2 + \sqrt{9-8(2k^2-1)^2}\right] , 
\nonumber \\
b &=& \textstyle{1 \over 6} \left[1 + 4k^2 - \sqrt{9-8(2k^2-1)^2}\right] .
\end{eqnarray}
Furthermore $K \equiv K(k)$ and $\Pi(x,k)$ for $x = a,b$ denote the
complete elliptic integrals of the first and the third kind,
respectively. The angle $\alpha$ traverses the interval $[0,\pi]$ by
decreasing $k^2$ from $k^2 = 1/2$ to $k^2 = 0$ in \Eq{P1g1}, changing
to \Eqs{P2g2}{ab} at $k^2 = 0$ and increasing $k^2$ back to $k^2 =
1/2$. The special point $k^2 = 0$ has no particular physical
significance, it only marks a singular point in the above parametric
representation of $g(\alpha)$. From \Eq{P2g2} we identify $k^2 =
(2-\sqrt{3})/4$ as the parameter value, where $g(\alpha) = 0$ or,
equivalently, $\alpha = \pi/3$.

\setlength{\columnwidth}{14.0cm}
\begin{figure}[t]
\centerline{\epsfbox{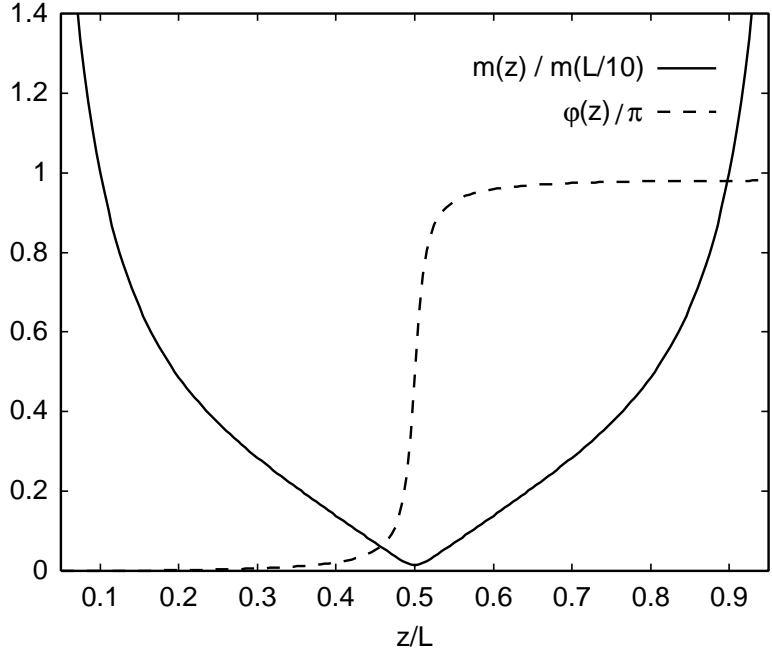}}
\centerline{\caption{\protect\small{
Amplitude $m(z)$ (solid line) and phase $\varphi(z)$ (dashed line) of
a two component order parameter for surface fields at an angle of
$\alpha = 0.98 \pi$ [$k^2 \simeq 0.499$, see
\protect\Eqs{alpc}{P2g2} and main text]. $m(z)$ has
been normalized to $m(L/10)$ so that $m(z)$ and $\varphi(z)$ can be
plotted on the same scale.
\label{mphiplot}}}}
\end{figure}
\setlength{\columnwidth}{\oldcolwidth}
Setting $k^2 = 1/2$ in \Eq{P1g1} yields $m(z) = m_{+,+}(z)$ [see
\Eq{mpp} for $\tau = 0$], $g(\alpha=0) = f_{+,+}(y=0)$ [see
\Eqs{fpp}{fab}], and $\varphi(z) = 0$ [see \Eq{alpc}]. This
means that the order parameter profile is given by ${\bf M}(z) =
\sqrt{12/g}\ (m_{+,+}(z),0,\dots ,0)$ as anticipated from the case
$N = 1$ for $(+,+)$ boundary conditions. In the limit $k^2 \to 1/2$
\Eqs{P2g2}{ab} yield $m(z) = |m_{+,-}(z)|$ [see \Eq{mpm} for $\tau =
0$], $g(\alpha = \pi) = f_{+,-}(y=0)$ [see \Eqs{fpm}{fab}], whereas
$\varphi(z)$ here is given by the {\em step function} $\varphi(z) =
\pi\ \theta(z/L - 1/2)$. The order parameter profile is then given by
${\bf M}(z) = \sqrt{12/g} (m_{+,-}(z),0,\dots ,0)$ which shows that
also for {\em antiparallel} surface fields mean field theory for an
$N$-component order parameter is already captured by the case $N = 1$.
We illustrate this remarkable behavior of ${\bf M}(z)$ for $\alpha /
\pi = 0.98$, i.e., a situation close to antiparallel surface fields.
The corresponding modulus $k$ [see \Eqs{P2g2}{ab}] is given by $k^2
\simeq 0.499$. The phase $\varphi(z)$ and the amplitude $m(z)$ of the
order parameter are shown in Fig. \ref{mphiplot}. The order parameter
rotates by almost the full amount $\alpha$ in a narrow
interval around $z = L/2$, where $m(z)$ is smallest. In the limit
$\alpha \to \pi$ this interval shrinks to the point $z = L/2$, where
$m(z)$ vanishes and $\varphi(z)$ becomes discontinuous.

Although the Casimir force is governed by universal scaling functions
\cite{MKSD92,MNB83} it is not possible to estimate their absolute
magnitude within Landau (mean field) theory. The reason is that for the
boundary conditions considered here these scaling functions contain a
common prefactor which depends on the bare coupling constant $g$ and
therefore has a value inaccessible by pure mean field arguments. In
order to at least partly fill this gap we now turn to the field
theoretic analysis of the Casimir force {\em at} bulk criticality.

\section{Casimir amplitudes}
At the bulk critical temperature $T = T_c$ $(\tau = 0)$ the Casimir
forces in a film are governed by the universal Casimir amplitudes
$\Delta_{a,b}$ which explicitly depend on the two surface universality
classes combined in the film. For $N > 1$ order parameter components
the Casimir amplitudes may also depend on continuously varying
parameters, as demonstrated above for the case $N = 2$ with tilted
surface fields. For $T \neq T_c$ $(\tau \neq 0)$ these amplitudes have
to be replaced by universal scaling functions $\theta_{a,b}(y)$ of a
suitably chosen scaling argument $y$ \cite{MKSD92}, which will not be
considered in this section.

The Casimir amplitude is defined as the finite-size amplitude of the
{\em free energy} of a film at bulk criticality \cite{MNB83,MKSD92}.
Translating this definition to the {\em force} one finds
\begin{equation}
\label{TppDel}
-{\partial \over \partial L} f(T_c,L) =
\langle T_{\perp \perp} \rangle = (d - 1) \Delta_{a,b} L^{-d}
\end{equation}
in $d$ dimensions and for $\tau = 0$, where $f(T_c,L) \equiv {\cal F}
(T=T_c,L) / (A\ k_B T_c)$ is the critical part of the free energy per
unit area $A$ of the plates. Following Ref. \cite{MKSD92} $k_B
T_c$ is used as the natural energy unit for the free energy, where
$k_B$ is the Boltzmann constant. As a first step beyond Landau
theory the contribution of Gaussian fluctuations to the Casimir force,
i.e., to the amplitudes $\Delta_{a,b}$ will be investigated here. We
introduce the fluctuation part $\widetilde{\bf \Phi}$ of the order
parameter ${\bf \Phi}$ by ${\bf \Phi} = \overline{\bf \Phi} +
\widetilde{\bf \Phi}$, where $\overline{\bf \Phi} \equiv {\bf M}(z) =
\sqrt{12/g}\ (m(z),0,\dots ,0)$ is the mean field order parameter
profile discussed in the preceding section and Appendix A. Inserting
the above decomposition of ${\bf \Phi}$ into \Eqs{Hb}{Hs} for $\tau =
0$ and $c_1 = c_2 = 0$ and keeping only the quadratic terms in
$\widetilde{\bf \Phi} = (\widetilde{\phi}_1,\dots ,\widetilde{\phi}_N)$
we obtain 
\begin{eqnarray}
\label{HGauss}
{\cal H}[{\bf \Phi}] &=& {\cal H}[\overline{\bf \Phi}] + 
{1 \over 2} \int d^{d-1}x \int_0^L dz
\left\{(\nabla \widetilde{\bf \Phi})^2 +
\left[6m^2(z) \widetilde{\phi}_1^2 + 2m^2(z)
\left( \widetilde{\phi}_2^2 + \dots + \widetilde{\phi}_N^2 \right)
\right] \right\} \nonumber \\
&& + {\cal O}\left[\widetilde{\phi}_1 \widetilde{\bf \Phi}^2\right]
+ {\cal O}\left[(\widetilde{\bf \Phi}^2)^2\right] ,
\end{eqnarray}
where ${\bf h}_1 = \pm {\bf h}_2 = (h_1,0,\dots ,0)$ in the limit $h_1
\to \infty$ is implicitly assumed. The mean field contribution ${\cal
H}[\overline{\bf \Phi}]$ to \Eq{HGauss} has already been discussed in
\Eq{HPhi}. Following \Eq{HGauss} we decompose the Casimir force into
the mean field part, a Gaussian part, and higher order corrections
according to
\begin{equation}
\label{Tppdec}
\langle T_{\perp \perp} \rangle = \langle T_{\perp \perp} \rangle_0 +
\langle T_{\perp \perp} \rangle_1 + {\cal O}(g) =
(6/g) t_{\perp \perp} + \langle T_{\perp \perp} \rangle_1 + {\cal O}(g).
\end{equation}
In order to determine $\langle T_{\perp \perp} \rangle_1$ from
\Eq{Tkl} one also needs the cubic terms in \Eq{HGauss} and we will
therefore not follow this approach any further. It is much more
convenient to determine $\langle T_{\perp \perp} \rangle_1$ from the
Gaussian contribution to the free energy by taking its first
derivative with respect to the film thickness $L$ \cite{MKSD92}.
Following Ref. \cite{MKSD92} this can be done most easily in a
spectral representation of the Gaussian Hamiltonian given by
\Eq{HGauss}. For the evaluation of $\langle T_{\perp \perp} \rangle_1$
{\em only} the eigenvalue spectrum is needed. According to \Eq{HGauss}
the spectrum consists of a longitudinal part $\epsilon_n^{(2)}$
characterizing the eigenmodes of the longitudinal fluctuations
$\widetilde{\phi}_1$ of the order parameter and a transverse part
$\epsilon_n^{(1)}$ which is the same for each of the $N-1$ transverse
components $(\widetilde{\phi}_2,\dots ,\widetilde{\phi}_N)$ of the
order parameter fluctuations. The spectra $\epsilon_n^{(1)}$ and
$\epsilon_n^{(2)}$ are determined in Appendix B. Once the eigenvalues
are given one can employ the dimensional regularization scheme and
according to Ref. \cite{MKSD92} we find
\begin{equation}
\label{FGauss}
f(T_c,L) = {\cal H}[\overline{\bf \Phi}] +
{\Gamma\left[(3-d)/2\right] \over 2^{d-1} \pi^{(d-1)/2} (d-1)}
\left[ \sum_{n=3}^\infty \left(\epsilon_n^{(2)}\right)^{(d-1)/2} +
(N-1) \sum_{n=2}^\infty \left(\epsilon_n^{(1)}\right)^{(d-1)/2} \right]
\end{equation}
for the critical part free energy within the Gaussian approximation
in $d$ dimensions. The $L$ dependence of the Gaussian contribution to
$f(T_c,L)$ is completely determined by the $L$ dependence of the
eigenvalues. From simple dimensional analysis one has $\epsilon_n^{(i)}
\sim L^{-2}$ so that $d \epsilon_n^{(i)} / dL = -(2/L) \epsilon_n^{(i)}$
for $i=1,2$. From \Eqs{TppDel}{FGauss} we find
\begin{equation}
\label{TppGauss}
\langle T_{\perp \perp} \rangle = {6 \over g} t_{\perp \perp} +
{\Gamma\left[(3-d)/2\right] \over 2^{d-1} \pi^{(d-1)/2} L}
\left[ \sum_{n=3}^\infty \left(\epsilon_n^{(2)}\right)^{(d-1)/2} +
(N-1) \sum_{n=2}^\infty \left(\epsilon_n^{(1)}\right)^{(d-1)/2} \right]
+ {\cal O}(g)
\end{equation}
for the Casimir force in the Gaussian approximation.
The mode sums in \Eq{TppGauss} diverge for $d=4-\varepsilon$ and we
therefore employ the dimensional regularization scheme. Furthermore,
the above sums yield an UV singularity in the typical form
$1/\varepsilon$ which must be treated analytically in order to
facilitate the renomalization of \Eq{TppGauss}. Both objectives can be
achieved with the asymptotic expansions of the eigenvalues
$\epsilon_n^{(1)}$ and $\epsilon_n^{(2)}$ for large mode numbers $n$
which are given by \Eqs{specapp}{specapm}. The regularization of the
mode sums and the analytical treatment of the $1/\varepsilon$ pole is
summarized in Appendix C. Using the results from Appendix B and C we
now investigate the different boundary conditions separately, where
the mean-field results given by \Eqss{fpp}{fpm}{fmix} are only needed
for $k^2 = 1/2$ $(\tau = 0)$.

For the renormalization of the Casimir force given by \Eq{TppGauss} we
use the conventions of Ref. \cite{MKSD92} and define the renormalized
coupling constant $u$ by
\begin{equation}
\label{u}
g = 2^d \pi^{d/2} \mu^{4-d} Z_u u \quad , \quad
Z_u = 1 + {N + 8 \over 3}{u \over \varepsilon} + {\cal O}(u^2) ,
\end{equation}
where $\mu$ is an arbitrary momentum scale and $d = 4 - \varepsilon$
in the following. The infrared stable fixed point value
$u^*(\varepsilon)$ of the renormalized coupling constant $u$ is given
by \cite{HWD86}
\begin{equation}
\label{ufixN}
u^*(\varepsilon) = {3 \over N+8} \varepsilon + {9(3N+14) \over
(N+8)^3} \varepsilon^2 + {\cal O}(\varepsilon^3) .
\end{equation}
For later reference we also quote the 3-loop estimate \cite{KNSCL91}
\begin{equation}
\label{ufix1}
u_1^*(\varepsilon) = {\varepsilon \over 3} + {17 \over 81} \varepsilon^2
+ \left({709 \over 17496} - {4 \over 27}\ \zeta(3) \right)\varepsilon^3
+ {\cal O}(\varepsilon^4)
\end{equation}
of the fixed point value $u^*$ for $N = 1$, where $\zeta(3) \simeq
1.20206$ is a special value of the Riemann zeta function. In order to
improve the predictive quality of a low-order $\varepsilon$-expansion
for $\varepsilon = 1$ $(d = 3)$ in a simple way one may try to include
exact results for the quantity in question in $d = 2$ in the sprit of
a Pad{\'e} approximant in the variable $\varepsilon$. This can be
applied rather successfully to the Casimir amplitude $\Delta_{per}$
thus improving the agreement between the
field-theoretic prediction \cite{MKSD92} and the Monte-Carlo estimate
\cite{MKDPL96} in $d = 3$. We will therefore follow the same procedure
here, where the case of $(SB,+)$ boundary conditions must be excluded,
because the $SB$ multicritical point does not exist in $d = 2$.

The renormalized expression for $\langle T_{\perp \perp} \rangle$ for
$(+,+)$ boundary conditions can be obtained by inserting the mean field
result given by \Eq{fpp} for $y = 0$ and the regularized mode sum
given by \Eq{sumpp} into \Eq{TppGauss} and by applying the
renormalization prescription given by \Eq{u}. After expanding all $d$
dependent quantities to first order in $\varepsilon$ [see \Eq{G3d2}]
the $1/\varepsilon$ pole coming from \Eq{sumpp} is cancelled, i.e.,
the UV singularity has been consistently removed from the theory. The
Casimir force then follows by evaluating the resulting renormalized
expression for $\langle T_{\perp \perp} \rangle$ {\em at} the
renormalization group fixed point $u = u^*(\varepsilon)$ given by
\Eq{ufixN}. The $\varepsilon$-expansion of the universal Casimir
amplitude $\Delta_{+,+}$, which characterizes the strength of the
Casimir force in a critical film with parallel surface fields, is
finally obtained by applying the definition of $\Delta_{a,b}$ given by
\Eq{TppDel}. The algebraic manipulations involved here starting from
\Eqss{TppGauss}{sumpp}{u} are absolutely elementary, so that we only
quote the final result
\begin{equation}
\label{Dpp}
\Delta_{+,+} = -{\pi^{d/2} \Gamma(d/2) \over 2 u^*(\varepsilon)}
\left({K \over \pi}\right)^4 \left[1 - {9 \varepsilon \over N+8}\
0.6853 + \varepsilon\ {N-1 \over N+8}\  0.1242 +
{\cal O}(\varepsilon^2) \right] ,
\end{equation}
where part of the $\varepsilon$-expansion has been resummed
consistently to first order in $\varepsilon$ using \Eq{Gd2}.
For $(+,-)$ boundary conditions the same procedure can be applied
using \Eq{fpm} for $y = 0$ and \Eqs{TppGauss}{sumpm}. One finds
\begin{equation}
\label{Dpm}
\Delta_{+,-} = 2{\pi^{d/2} \Gamma(d/2) \over u^*(\varepsilon)}
\left({K \over \pi}\right)^4 \left[1 - {9 \varepsilon \over N+8}\
0.2822 + \varepsilon\ {N-1 \over N+8}\  0.4066 +
{\cal O}(\varepsilon^2) \right] .
\end{equation}
From \Eq{fmix} for $y = 0$ and \Eqs{TppGauss}{sumSBp} one has
for $(SB,+)$ boundary conditions
\begin{equation}
\label{DSBp}
\Delta_{SB,+} = -{\pi^{d/2} \Gamma(d/2) \over 32 u^*(\varepsilon)}
\left({K \over \pi}\right)^4 \left[1 + {9 \varepsilon \over N+8}\
1.7141 + \varepsilon\ {N-1 \over N+8}\  2.8448 +
{\cal O}(\varepsilon^2) \right] .
\end{equation}
From \Eq{fmix} for $y = 0$ and \Eqs{TppGauss}{sumOp} one finally has
for $(O,+)$ boundary conditions
\begin{equation}
\label{DOp}
\Delta_{O,+} = {\pi^{d/2} \Gamma(d/2) \over 8 u^*(\varepsilon)}
\left({K \over \pi}\right)^4 \left[1 + {9 \varepsilon \over N+8}\
0.1988 + \varepsilon\ {N-1 \over N+8}\  0.2289 +
{\cal O}(\varepsilon^2) \right] .
\end{equation}
Note that $u^*(\varepsilon)$ in the above expressions is given by
\Eq{ufixN}. It is remarkable that the coefficients of the Gaussian
contribution to $\Delta_{SB,+}$ given by \Eq{DSBp} are much bigger
than the corresponding coefficients in \Eqss{Dpp}{Dpm}{DOp}. This may
be due to the fact that the order parameter near a $SB$ surface is
much more susceptible to fluctuations than near $O$ or $E$ surfaces.

In the Ising universality class $(N = 1)$ in $d = 2$ three of the
above Casimir amplitudes are known exactly from conformal field theory
\cite{JLC86}. They are given by
\begin{equation}
\label{D2d}
\Delta_{+,+} = -{\pi \over 48} \quad , \quad
\Delta_{+,-} = {23 \over 48} \pi \quad , \quad
\Delta_{O,+} = {\pi \over 24} .
\end{equation}
The construction of a Pad{\'e} approximant from \Eqss{Dpp}{Dpm}{DOp}
for $N = 1$ which extrapolates to the amplitudes given by \Eq{D2d} for
$\varepsilon = 2$ is arbitrary to a certain degree. If one uses
\Eq{ufix1} instead of \Eq{ufixN} for $N = 1$ and introduces an
additional $\varepsilon^2$-contribution to the square bracket of
\Eqss{Dpp}{Dpm}{DOp} such that \Eq{D2d} is reproduced for $\varepsilon
= 2$ one finds the interpolation formulas
\begin{eqnarray}
\label{DPade}
\Delta_{+,+} &=& -{\pi^{d/2} \Gamma(d/2) \over 2 u_1^*(\varepsilon)}
\left({K \over \pi}\right)^4 \left[1 - 0.6853\ \varepsilon + 0.1275\
\varepsilon^2 \right] , \nonumber \\ \nonumber \\
\Delta_{+,-} &=& 2{\pi^{d/2} \Gamma(d/2) \over u_1^*(\varepsilon)}
\left({K \over \pi}\right)^4 \left[1 - 0.2822\ \varepsilon + 0.0914\
\varepsilon^2 \right] , \\ \nonumber \\
\Delta_{O,+} &=& {\pi^{d/2} \Gamma(d/2) \over 8 u_1^*(\varepsilon)}
\left({K \over \pi}\right)^4 \left[1 + 0.1988\ \varepsilon - 0.0707\
\varepsilon^2 \right] . \nonumber
\end{eqnarray}
Numerical estimates of the Casimir amplitudes in $d = 3$ obtained from
the above analytical formulas are summarized in Table I.
\begin{center}
\parbox{14.7cm}{\protect\small{Table I.
Casimir amplitudes for the Ising universality class in $d = 3$. The
values labelled by $\varepsilon = 1$ are obtained by evaluating
Eqs. (\protect\ref{DperOO}), (\protect\ref{Dpp}), (\protect\ref{Dpm}),
(\protect\ref{DSBp}), and (\protect\ref{DOp}) for $N=1$ and
$\varepsilon=1$. The values labelled $d=3$ are obtained from
\protect\Eqs{DperOPade}{DPade} for $d = 3$ $(\varepsilon=1)$. The
Monte-Carlo estimates obtained from the serial version of the
algorithm presented in Ref. \protect\cite{MKDPL96} are labelled by
'MC' (see also main text). Statistical errors (one standard deviation)
are in the last two digits as indicated inside the parenthesis. The
last line shows Migdal-Kadanoff estimates taken from Ref.
\protect\cite{INW86}.}}
\end{center}
\begin{center}
\begin{tabular}{lllllll}
\hline \hline
& $\ \ \Delta_{per}$ & $\ \ \Delta_{O,O}$ & $\ \ \Delta_{+,+}$ &
$\ \Delta_{+,-}$ & $\ \Delta_{SB,+}$ & $\ \Delta_{O,+}$ \\
\hline
$\varepsilon = 1$ & $-0.1116$ & $-0.0139$ & $-0.173$ & $1.58$ &
$-0.093\ \ \ $ & $0.165$ \\
$d = 3$ & $-0.1315$ & $-0.0164$ & $-0.326$ & $2.39$ & & $0.208$ \\
MC & $-0.1526(10)\ $ & $-0.0114(20)\ $ & $-0.345(16)\ $ &
$2.450(32)\ $ & & $0.1873(70)\ $ \\
Ref.\cite{INW86} & & $-0.015$ & $\ \ 0$ & $0.279$ & $\ \ \ 0.017$ &
$0.051$ \\
\hline \hline
\end{tabular}
\end{center}

In $d = 3$ and for $N = 1$ the Casimir amplitudes $\Delta_{O,O}$,
$\Delta_{O,+}$, $\Delta_{+,+}$, and $\Delta_{+,-}$ can be measured by
a Monte-Carlo simulation of the Ising model defined by \Eq{HIsing}.
The algorithm and its special adaptation to the measurement of the
Casimir amplitude is presented in Ref. \cite{MKDPL96} in detail. We
therefore only briefly describe the differences between the
implementations used here and in Ref. \cite{MKDPL96}. The present
implementation of the algorithm utilizes a {\em serial} hybrid update
scheme which consists of a Metropolis update sweep of the whole lattice
followed by a Wolff update. The length of the equilibration and the
measurement period used here correspond to those in Ref. \cite{MKDPL96}.
The slab geometry contains $L'^2 \times L$ lattice sites, where $L'/L$
must be chosen as large as possible in order to approximate the
infinite slab geometry. In practice already $L' = 4L$ turns out to be
sufficient, i.e., the results obtained for this choice agree with
those for $L' = 6L$ within a fraction of one standard deviation. The
thickness $L$ of the slab has been varied between $L = 12$ and $L =
32$ layers. As in Ref. \cite{MKDPL96} we use the multiple histogram
technique \cite{AMFRHS89}, where the number of histograms taken has
been increased from 25 to 31 for $L > 24$ on order to guarantee
sufficient overlap between adjacent histograms \cite{MKDPL96}. The
simulations were run on DEC Alpha workstations at the University of
Wuppertal and the total amount of CPU time used is equvalent to about
one year of CPU time on a DEC 3000 workstation.

\begin{figure}[t]
\centerline{\epsfbox{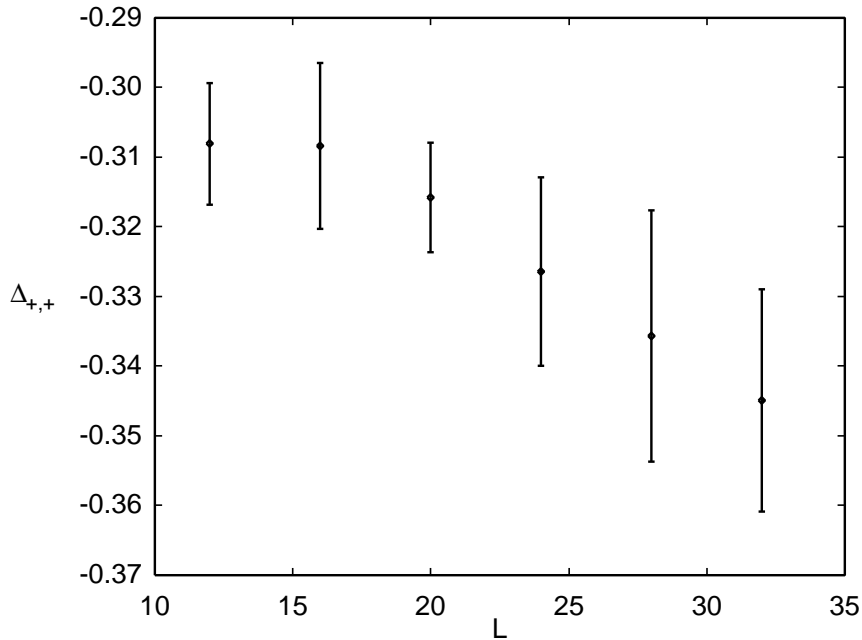}}
\centerline{\caption{\protect\small{
Monte-Carlo estimates of the Casimir amplitude $\Delta_{+,+}$ as a
function of the number of layers $L$ in a $L'^2 \times L$ slab for $L'
= 4L$. The size of the error bars represents one standard deviation.
The data point at $L = 32$ is taken as the final estimate (see Table I).
\label{Dppplot}}}}
\end{figure}

\begin{figure}[t]
\centerline{\epsfbox{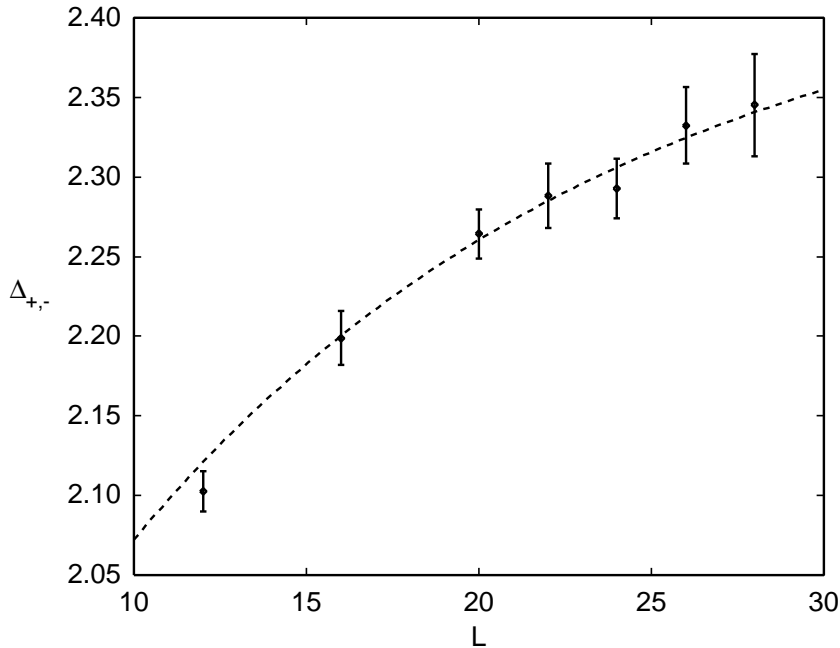}}
\centerline{\caption{\protect\small{
Monte-Carlo estimates of the Casimir amplitude $\Delta_{+,-}$ as a
function of the number of layers $L$ in a $L'^2 \times L$ slab for $L'
= 4L$. The size of the error bars represents one standard deviation.
The dashed line shows a fit of \protect\Eq{Dpmeff} to the data for $L
\geq 16$ giving the estimate of $\Delta_{+,-}$ displayed in Table I.
The absolute size of the errorbars is about
twice that in Fig. \protect\ref{Dppplot}.
\label{Dpmplot}}}}
\end{figure}

The serial implementation of the algorithm has been tested for the
Casimir amplitude $\Delta_{per}$ with $L' = 4L$ and $L' = 6L$ for $L
=20$ and $L = 24$. The estimates for $\Delta_{per}$ obtained with
these four lattice sizes agree within their statistical error and give
the final estimate
\begin{equation}
\label{DperMC}
\Delta_{per} = -0.1526 \pm 0.0010 ,
\end{equation}
which is in perfect greement with the estimate obtained from the
parallel algorithm \cite{MKDPL96}. The amplitude $\Delta_{O,O}$ has
been measured for the same lattices sizes and for $\Delta_{O,+}$
additional simulations were performed with $L = 28$ and $L' = 4L$. All
individual measurements agree within their statistical error and the
final estimates are shown in Table I. For $(+,+)$ and
$(+,-)$ boundary conditions, however, the situation is different. For
$\Delta_{+,+}$ measurements have been made for $12 \leq L \leq 32$ and
$L' = 4L$, the individual estimates are displayed in Fig.\ref{Dppplot}
as a function of $L$.
\setlength{\columnwidth}{14.0cm}
\setlength{\columnwidth}{\oldcolwidth}
The estimates show a clear
systematic dependence on $L$ and apparently even for $L = 32$ layers
the asymptotic regime has not yet been reached. The last three data
points fall onto a straight line within their error bars so that the
data cannot be extrapolated to an asymptotic value. As the current
Monte-Carlo estimate for $\Delta_{+,+}$ we therefore take the
measurement for the biggest system ($L = 32$, $L' = 4L$) (see Table I).
The situation for $\Delta_{+,-}$ is similar. The
individual measurements are shown in Fig. \ref{Dpmplot} for $12 \leq L
\leq 28$ $(L' = 4L)$. Again, the asymptotic regime has not been
reached for the biggest system, but this time it is possible to
estimate the asymptotic value for $\Delta_{+,-}$ by a least square fit
of the function
\begin{equation}
\label{Dpmeff}
\Delta_{+,-}^{eff}(L) = \Delta_{+,-} + D \exp(-\kappa L)
\end{equation}
to the data for $L \geq 16$ using $\Delta_{+,-}$, $D$, and $\kappa$ as
fit parameters.
\setlength{\columnwidth}{14.0cm}
\setlength{\columnwidth}{\oldcolwidth}
The exponential $L$ dependence of
$\Delta_{+,-}^{eff}(L)$ in \Eq{Dpmeff} is motivated by the
short-ranged nature of the interaction in \Eq{HIsing}. The error of
the amplitude $\Delta_{+,-}$ is estimated by taking the maximal
error of the individual measurements involved in the fit. All
estimates obtained from \Eqss{Dpp}{Dpm}{DOp} for $N = 1$, from \Eq{DPade},
and our Monte-Carlo estimates are summarized in Table I. For
completeness we also display estimates for $\Delta_{per}$
and $\Delta_{O,O}$ obtained from the partially resummed
$\varepsilon$-expansions \cite{MK94}
\begin{equation}
\label{DperOO}
\Delta_{per} = -N{\Gamma(d/2) \zeta(d) \over \pi^{d/2}}\left(1 - {5
\over 4} {N+2 \over N+8}\ \varepsilon + {\cal O}(\varepsilon^2) \right)
\quad , \quad
\Delta_{O,O} = 2^{-d} \Delta_{per}
\end{equation}
for $N = 1$ and from the Pad{\'e} approximants \cite{MK94,MKDPL96}
\begin{equation}
\label{DperOPade}
\Delta_{per} = -{\Gamma(d/2) \zeta(d) \over \pi^{d/2}}\left(1 - {5
\over 4} {4-d \over 7-d} \right) \quad , \quad
\Delta_{O,O} = 2^{-d} \Delta_{per} ,
\end{equation}
which reproduce the exact results \cite{JLC86}
\begin{equation}
\label{DperO2d}
\Delta_{per} = -{\pi \over 12} \quad , \quad
\Delta_{O,O} = -{\pi \over 48}
\end{equation}
in $d = 2$. For comparison we also reproduce Migdal-Kadanoff estimates
for the Casimir amplitudes in $d = 3$ from Ref. \cite{INW86}. The
agreement between the Pad{\'e} approximants and the Monte-Carlo
estimates is quite satisfactory, except for $\Delta_{O,O}$ which seems
to be closer to the partially resummed $\varepsilon$-expansion and the
Migdal-Kadanoff estimate. However, the amplitude is rather small and
therefore the relative statistical error of the Monte-Carlo estimate,
which is one standard deviation, is very large (20\%, see Table I).
In view of Fig. \ref{Dppplot} the Monte-Carlo estimate
for $\Delta_{+,+}$ given in Table I constitutes only an
upper bound for the true amplitude and must therefore also be handled
with caution. The fit procedure used to extract $\Delta_{+,-}$ from
the data shown in Fig. \ref{Dpmplot} is also susceptible to systematic
errors to a certain extent. However, compared to the parameters $D$
and $\kappa$ in \Eq{Dpmeff} the resulting estimate for $\Delta_{+,-}$
is quite robust with respect to, e.g., changes in the number of data
points included in the fit. The obtained variation of $\Delta_{+,-}$
is in the same order of magnitude as the statistical error given in
Table I. With regard to their reliability the analytical
and the Monte-Carlo estimates of $\Delta_{+,+}$, $\Delta_{+,-}$, and
$\Delta_{O,+}$ seem to be a substantial improvement over the
Migdal-Kadanoff results.

\section{Experimental implications}
A typical experimental setting, within which the film geometry
considered here is of particular interest, is provided by wetting
experiments performed on plane and chemically homogeneous substrates
\cite{SD88,MPNJOI85,MKSD92a}. The equilibrium thickness $L$ of the
wetting layer is determined by the {\em minimum} of the effective
interface potential \cite{SD88}. It is given by the grand canonical
free energy of a liquid layer of a {\em prescribed} thickness $l$,
which is in contact with the substrate on one side and with the bulk
vapor phase on the other side. In the limit of large interfacial areas
$A$ the effective interface potential can be written in the form
\cite{SD88,MKSD92a,SDAL89}
\begin{equation}
\label{oml}
\lim_{A \to \infty} {\Omega(T,l) \over A} \equiv \omega(l) = l\
\left[\rho_l(T)/\rho_v(T) - 1\right] p_0(T)\ \delta p + \sigma_{sl}(T)
+ \sigma_{lv}(T) + \delta \omega(T,l) ,
\end{equation}
where $\rho_l(T)$ and $\rho_v(T)$ are the liquid and the vapor
density, respectively and $p_0(T)$ denotes the liquid-vapor
coexistence line in a $p,T$ phase diagram. The quantity $\delta p$ in
\Eq{oml} is a dimensionless measure of the undersaturation of the
vapor, i.e., $\delta p > 0$ indicates that in the bulk the {\em vapor}
phase is thermodynamically stable. The substrate-liquid and
liquid-vapor interfacial tensions $\sigma_{sl}(T)$ and
$\sigma_{lv}(T)$ do not depend on $l$ and $\delta \omega(T,l)$
contains the dispersion (van der Waals) forces and the critical
Casimir forces in the liquid layer. For a binary liquid mixture as the
wetting agent the critical point of interest is the critical end point
of the line of critical demixing transitions on the liquid-vapor
coexistence surface (see Fig. 1 in Ref. \cite{SDAL89}). In order to
discuss the effect of criticality on the equilibrium thickness $L$ of
the wetting layer \cite{MPNJOI85,MKSD92a} we assume in the following
that the critical temperature $T_{cep}$ associated with this
critical end point is located {\em above} the wetting temperature
$T_w$ so that the condition $T_w < T \simeq T_{cep}$ guarantees a {\em
macroscopic} wetting layer of a {\em critical} binary liquid mixture. For
large values of $l$ the van der Waals contribution to $\delta
\omega(T,l)$ has the asymptotic form \cite{ESSCHA73}
\begin{equation}
\label{domvdW}
\delta \omega_{vdW}(T,l) = \left\{
\begin{array}{ll}
W(T)\,l^{-2} + {\cal O}(l^{-3}) & \mbox{(nonretarded)} \\
W_r(T)\,l^{-3} + {\cal O}(l^{-4}) & \mbox{(retarded)}.
\end{array} \right.
\end{equation}
The explicit temperature dependence of the Hamaker constant $W(T)$ and
its retarded counterpart $W_r(T)$ is quite weak and can be disregarded
in the critical regime around $T_{cep}$. According to \Eq{oml} one has
with $\delta \omega(T,l) = \delta \omega_{vdW}(T,l)$ taken from
\Eq{domvdW} $L(\delta p) \propto (\delta p)^{-1/3}$ in the nonretarded
case and $L(\delta p) \propto (\delta p)^{-1/4}$ in the retarded case.
Provided, the the wetting layer becomes thick enough, one observes a
crossover from the former to the latter power law for $\delta p \to 0$
in a wetting experiment, because the van der Waals forces become
retarded as $L$ increases \cite{ESSCHA73}. At the critical end point
$\delta \omega$ is modified by the long-ranged Casimir forces
according to
\begin{equation}
\label{domc}
\delta \omega(T_{cep},l) = \delta \omega_{vdW}(T_{cep},l) + k_B
T_{cep} \Delta_{a,b} l^{-(d-1)}
\end{equation}
in $d$ dimensions, where $k_B$ is the Boltzmann constant and
$\Delta_{a,b}$ is the Casimir amplitude for boundary conditions of
type $(a,b)$ as discussed in the preceding section. If the van der
Waals forces are not retarded one can combine \Eqs{domvdW}{domc} in $d
= 3$ by defining the effective Hamaker constant \cite{MKSD92a}
\begin{equation}
\label{Weff}
W_{eff} \equiv W + k_B T_{cep} \Delta_{a,b},
\end{equation}
where the temperature dependence of $W$ has been disregarded. The
effective Hamaker constant $W_{eff}$ replaces $W$ in the
effective interface potential given by \Eq{oml} and thus determines
the equilibrium thickness $L$ of the wetting layer for fixed
undersaturation $\delta p$. The ratio $R(\delta p)$ of the wetting
layer thickness $L_{cep}(\delta p)$ {\em at} the critical end point
and the thickness $L(\delta p)$ of the wetting layer {\em outside} the
critical regime is then determined by the ratio $W_{eff}/W$
\cite{MKSD92a}. One obtains
\begin{equation}
\label{LcL}
R(\delta p) \equiv L_{cep}(\delta p) / L(\delta p)
= (W_{eff}/W)^{1/3} = (1 + k_B T_{cep} \Delta_{a,b} / W)^{1/3} ,
\end{equation}
which is independent of the undersaturation $\delta p$ to leading
order in $\delta p$ (see Ref. \cite{MKSD92a} for details). If both the
liquid-substrate and the liquid-vapor interface prefer the same
component of the binary liquid mixture one has $(a,b) = (+,+)$ and
\Eq{LcL} predicts a thinning of the wetting layer, because
$\Delta_{+,+} < 0$ (see Table I). In the opposite case
$(a,b) = (+,-)$ applies and \Eq{LcL} predicts an increase in the
wetting layer thickness due to $\Delta_{+,-} > 0$. An experimental
realization for the latter case is provided by a methanol-hexane
mixture on Si - SiO$_2$ wafers as substrates \cite{AMBML96}. The
mixture wets the wafers at a temperature below $T_{cep} \simeq 300K$,
where the methanol concentration is enhanced near the substrate and
the hexane concentration is enhanced near the liquid-vapor interface
providing a realization of the $(+,-)$ boundary condition. The Hamaker
constant for this system is given by $W \simeq 9 \times 10^{-15}$erg
\cite{AMBML96} and with $\Delta_{+,-} \simeq 2.4$ taken from Table I
one obtains $R(\delta p \to 0) \simeq 2.3$ from \Eq{LcL}.
The corresponding value of $R$ for $^4$He on Ne substrates at the
lower $\lambda$-point is $R \simeq 0.995$ \cite{MKSD92a}. The
explanation for this drastic difference is twofold. First, there is
the combined effect of the Hamaker constant $W$ and the relevant
energy scale given by $k_B T_c$. For methanol-hexane on Si - SiO$_2$
one has $T_c = T_{cep} \simeq 300K$ so that $W/(k_B T_{cep}) \simeq
0.2$, whereas for $^4$He on Ne one has $T_c = T_\lambda = 2.17K$ which
implies $W/(k_B T_\lambda) \simeq 2$ \cite{MKSD92a}. Second, the
relavant Casimir amplitude is $\Delta_{+,-} \simeq 2.4$ for methanol
hexane and $\Delta_{O,O} \simeq -0.022$ for $^4$He \cite{MKSD92a}. In
the ratio $(W_{eff}-W)/W$ [see \Eqs{Weff}{LcL}] one therefore has one
factor $\sim 10$ in favor of methanol-hexane coming from $W/(k_B
T_c)$ and a second factor $\sim 100$ in favor of methanol-hexane
from the Casimir amplitude which combine to the observed drastic
quantitative difference in $R(\delta p)$.

For $\delta p \to 0$ the equilibrium thickness $L(\delta p)$ of the
wetting layer increases so that the van der Waals forces may become
retarded [see \Eq{domvdW}]. In the retarded regime the critical
contribution to $\delta \omega(T_{cep},l)$ becomes the leading term in
\Eq{domc} for $d = 3$ and therefore $R(\delta p)$ defined by \Eq{LcL}
{\em diverges} for $\delta p \to 0$ according to \cite{MKSD92a}
\begin{equation}
\label{Rret}
R(\delta p \to 0) = \left({2 k_B T_{cep} \Delta_{+,-} \over \rho_l -
\rho_v}\right)^{1/3} \left({3 W_r \over \rho_l - \rho_v}\right)^{-1/4}
\left({p_0 \over \rho_v}\right)^{-1/12} (\delta p)^{-1/12} .
\end{equation}
For $(+,+)$ boundary conditions one has $\Delta_{+,+} < 0$ and in this
case retardation of the van der Waals forces leads to a {\em finite}
value of $L_{cep}(\delta p)$ for $\delta p \to 0$. The ratio $R(\delta
p)$ then {\em vanishes} as
\begin{equation}
\label{Rfin}
R(\delta p \to 0) = {\rho_l - \rho_v \over -2k_B T_{cep} \Delta_{+,+}}
\left({3 W_r \over \rho_l - \rho_v}\right)^{3/4} \left({p_0 \over
\rho_v}\right)^{1/4} (\delta p)^{1/4}
\end{equation}
for $\delta p \to 0$ \cite{MKSD92a}. The amplitudes of the power laws
governing $R(\delta p \to 0)$, which according to \Eqs{Rret}{Rfin}
depend on the product $k_B T_{cep} \Delta_{a,b}$, show the same
sensitivity to the type of the wetting agent (methanol-hexane or
$^4$He) as the effective Hamaker constant (see above). The drastic
enhancement of $k_B T_{cep} \Delta_{a,b}$ observed for typical binary
liquid mixtures in comparison with $^4$He makes critical effects on
wetting layers much easier to detect experimentally. A corresponding
statement can be made for direct force measurements by atomic force
microscopes \cite{JNIPMM88}. If two parallel plates at distance $L$
are immersed into a binary liquid mixture, which is close to its
critical demixing transition, the force per unit area $K_c$ between
the plates will deviate from the bulk pressure due to the {\em finite}
distance between the plates. This deviation is given by \cite{MKSD92a}
\begin{equation}
\label{dKc}
\delta K_c(L) = K_c(L) - K_c(L = \infty) = -{\partial
\over \partial L} \delta \omega(T_c,L) = 2 W_{eff} L^{-3}
\end{equation}
if the van der Waals forces are not retarded [see \Eqs{domc}{Weff}].
Note that $T_c$ in \Eq{dKc} is {\em not} given by $T_{cep}$. Here
$T_c$ marks a second order phase transition from the demixed to the
mixed liquid, which takes place {\em inside} the liquid regime in the
phase diagram away from the liquid-vapor coexistence surface (see Fig.
1 in Ref. \cite{SDAL89}). However, typically $T_c$ is roughly about
the same size as $T_{cep}$. By inserting the values for
$\Delta_{+,+}$ and $\Delta_{+,-}$ (see Table I), $T_c
\simeq 300K$, and $W \simeq 9 \times 10^{-15}$erg for methanol-hexane
into \Eq{Weff} one finds
\begin{equation}
\label{Weffeval}
W_{eff} / W \simeq \left\{
\begin{array}{rl}
-0.6 & \mbox{for $(+,+)$ boundary cond.} \\
12 & \mbox{for $(+,-)$ boundary cond.}
\end{array} \right. .
\end{equation}
According to \Eq{Weffeval} the critical contribution to $\delta
K_c(L)$ can lead to a sign reversal of $\delta K_c(L)$ for equal
plates and increases $\delta K_c(L)$ by an order of magnitude for
opposing plates. The effects of criticality on $\delta K_c(L)$ should
therefore be detectable by direct force measurements in critical
binary liquid mixtures.

\section{Summary and discussion}
If macroscopic bodies are immersed in a critical fluid long-ranged
forces between these bodies are generated by critical fluctuations
of the order parameter. For the special case of binary liquid mixtures
confined to a parallel plate geometry these forces have been analyzed
for various boundary conditions involving surface fields in order to
describe chemical affinities of the confining walls or interfaces
towards one of the components of the mixture. In particular, the
following results have been obtained:

1. Within mean-field (Landau) theory for an Ising-like system ($N = 1$
order parameter components) the universal scaling functions $f_{+,+}(y)$
and $f_{+,-}(y)$ of the Casimir force can be easily obtained in a
parameter representation without detailed knowledge about the order
parameter profile. Either scaling function indicates that the
corresponding Casimir forces should be visible over a surprisingly
broad range in the scaling variable $y = \tau L^{1/\nu}$. The scaling
functions $f_{SB,+}(y)$ and $f_{O,+}(y)$ can be obtained from
$f_{+,+}(y)$ and $f_{+,-}(y)$ by applying a simple scale
transformation to $f$ and $y$. In comparison with $(+,+)$ and $(+,-)$
boundary conditions the Casimir forces for these mixed boundary
conditions are substantially reduced both in their magnitude and in
the range of the scaling argument  $y$ over which they are visible.
For $(+,+)$ and $(SB,+)$ boundary conditions the force is attractive,
for $(+,-)$ and $(O,+)$ boundary conditions it is repulsive. For $N
\geq 2$ an additional degree of freedom in the choice of the boundary
conditions (surface fields) is provided by the introduction of an
arbitrary tilt angle $\alpha$ between the surface fields. For $N = 2$
order parameter components and $y = 0$ it is shown that the amplitude
function $g(\alpha)$ smoothly interpolates between the special values
(Casimir amplitudes) $f_{+,+}(0)$ $(\alpha = 0)$ and $f_{+,-}(0)$
$(\alpha = \pi)$ of the scaling functions. The Casimir force vanishes
for $\alpha = \pi/3$. For $\alpha = \pi$ the order parameter profile
is identical to the profile for $N = 1$ and $(+,-)$ boundary
conditions. For critical binary liquid mixtures only the case $N = 1$
is relevant.

2. For the special case $y = 0$ $(T = T_{c,bulk})$ the scaling
functions reduce to the universal Casimir amplitudes $\Delta_{a,b}$
for $(a,b)$ boundary conditions which have been calculated
analytically to one-loop order (Gaussian fluctuations) in order to
obtain quantitative estimates for the magnitude of the Casimir force
in $d = 3$. For the most relevant case $N = 1$ and for $(+,+)$,
$(+,-)$, and $(O,+)$ boundary conditions it is possible to contruct
Pad{\'e}-type approximants for the Casimir amplitudes in $d = 3$ by
including exact results from conformal field theory in $d = 2$ into an
interpolation scheme for the amplitudes as a function of $d$. If a
3-loop estimate for the fixed point value $u^*$ of the
renormalized coupling constant $u$ is used in the interpolation scheme
the resulting values for $\Delta_{+,+}$, $\Delta_{+,-}$, and
$\Delta_{O,+}$ in $d = 3$ agree quite well with corresponding
numerical esitimates from a Monte-Carlo simulation of an Ising model
confined to a slab geometry in $d = 3$ with surface fields. The
estimates indicate, that for a critical binary liquid mixture the
Casimir amplitudes are between one and two orders of magnitude larger
than the previously studied amplitude $\Delta_{O,O}$ for $^4$He at the
$\lambda$-transition.

3. For critical binary liquid mixtures confined between equal or
opposing walls the Casimir amplitudes $\Delta_{+,+}$ or $\Delta_{+,-}$,
respectively, yield the absolute strength of the Casimir force in
units of $k_B T_c$. The film geometry considered here is realized in a
natural way in the course of a wetting transition on a plane and
chemically homogeneous substrate. The special case of $(+,-)$ boundary
conditions is realized by the binary mixture methanol-hexane which
forms a macroscopic wetting layer on Si - SiO$_2$ wafers in the
vicinity of the critical end point of the demixing transitions.
Disregarding any temperature dependence of the Hamaker constant the
presence of critical fluctuations in the wetting layer leads to an
increase of the equilibrium layer thickness by more than a factor of
two. The corresponding critical effect on a wetting layer of $^4$He at
the lower $\lambda$-point is serveral orders of magnitude weaker. In
accordance with this observation critical fluctuations in binary
liquid mixtures have a strong effect on the effective Hamaker constant
which determines the strength of the force between two parallel plates
immersed into the mixture. Therefore, critical binary liquid mixtures
appear to be ideal candidates to probe the universal Casimir
amplitudes and the associated universal scaling functions by wetting
experiments or by direct force measurements using a suitably adapted
version of the atomic force microscope.

\acknowledgements
The author gratefully acknowledges useful correspondence with E.
Eisenriegler, B.M. Law, and A. Mukhopadhyay.

\appendix

\section{Order parameter profiles}
The order parameter profiles in a critical film within mean field
(Landau) theory for the Ginzburg - Landau Hamiltonian given by
\Eqs{Hb}{Hs} have already been discussed in the literature in some
detail for various reasons \cite{MEFHN81,HNMEF83,INW86,MIKANO72} (see
also Sec. I). Therefore we only summarize the main results of mean
field theory here for later reference. We restrict the analysis to
the case $N=1$ (Ising universality class). The Euler-Lagrange equation
for the order parameter profile $M(z)$ reads
\begin{equation}
\label{EL}
M''(z) = \tau M(z) + \textstyle{g \over 6} M^3(z) ,
\end{equation}
where the boundary conditions
\begin{eqnarray}
\label{ELbc}
M'(0) = c_1 M(0) - h_1 &\quad ,\quad& M'(L) = - c_2 M(L) + h_2
\end{eqnarray}
must be fulfilled. In order to obtain the leading asymptotic behavior
of $M(z)$ in the critical regime we only consider the limiting cases
$h_1 = h_2 \to \infty$ [$(+,+)$ boundary conditions] and $h_1 = -h_2
\to \infty$ [$(+,-)$ boundary conditions] in \Eq{ELbc}. In this limit
the order parameter profile has the singularities $M(z) \sim 1/z$
for $z \to 0$ and $M(z) \sim 1/(L-z)$ for $z \to L$. This singularity
of $M(z)$ at the system boundaries just constitutes the mean field
description of the asymptotic increase $\overline{\Phi}(z) \sim
z^{-\beta/\nu}$ of the order parameter profile as $z \to 0$ for large
(or infinite) surface fields. For this asymptotic power law
to be valid the condition $a \ll z \ll \xi$ must be fulfilled, where
$a$ is a typical {\em microscopic} length scale and $\xi$ is the
correlation length. In a lattice model for example $a$ is given by
the lattice constant. The order parameter profile for such a model
will deviate from this power law increase on the scale $z \sim a$ away
from the surface and take a {\em finite} value right at the surface
even for an infinite surface field.

In order to simplify the notation for the following considerations we
introduce the order parameter function $m(z)$ by setting $M(z) =
\sqrt{12/g}\ m(z)$ in \Eq{EL}, where $m(z)$ solves the modified
Euler-Lagrange equation
\begin{equation}
\label{ELm}
m''(z) = \tau m(z) + 2 m^3(z) .
\end{equation}
We furthermore suppress the parametric dependence of $m(z)$ on the reduced
temperature $\tau$ in the notation. Multiplying \Eq{ELm} by $m'(z)$
one finds
\begin{equation}
\label{ELint}
m'^2(z) = \tau\ m^2(z) + m^4(z) + m'^2(z_0) - \tau\ m^2(z_0) - m^4(z_0)
\end{equation}
as the first intgral of \Eq{ELm}, where $z_0$ is an arbitrary reference
point $0 < z_0 < L$. For the combinations $(+,+)$ and $(+,-)$ of boundary
conditions considered here $z_0 = L/2$ is a convenient choice, because
m(z) is either a symmetric or an antisymmetric function with respect
to the midplane $z = L/2$, respectively (see also Refs.
\cite{AOP92,KBREDPLAMF96,KBDPLAMF95}). Up to an overall factor the
integration constant in \Eq{ELint} can be identified with $\langle
T_{\perp \perp} \rangle$ in the mean field approximation which we
denote by $\langle T_{\perp \perp}\rangle_0$ [see also \Eq{Tkl}]. We
define $\langle T_{\perp \perp} \rangle_0 \equiv (6/g)\ t_{\perp
\perp}$, so that
\begin{equation}
\label{tpp}
t_{\perp \perp} = m'^2(L/2) - \tau\ m^2(L/2) - m^4(L/2)
\end{equation}
is just the integration constant on the r.h.s. of \Eq{ELint}. With the
substitution $m^2(z) \equiv P(z) - \tau/3$ \Eq{ELint} takes the
form
\begin{equation}
\label{P}
P'^2(z) = 4[P(z)-e_1][P(z)-e_2][P(z)-e_3] ,
\end{equation}
where
\begin{equation}
\label{e1e2e3}
e_1 = -\tau/6 + \sqrt{\tau^2/4 - t_{\perp \perp}}\ ,\ 
e_2 = \tau/3\ ,\ 
e_3 = -\tau/6 - \sqrt{\tau^2/4 - t_{\perp \perp}} .
\end{equation}
From the obvious property $e_1 + e_2 + e_3 = 0$ and
the structure of \Eq{P} it is immediately clear that $P(z) = m^2(z) +
\tau/3$ is given by a Weierstrass elliptic function $\wp(z;g_2,g_3)$
with the invariants
\begin{eqnarray}
\label{g2g3}
g_2 &=& -4(e_1 e_2 + e_2 e_3 + e_3 e_1) = 4(\tau^2/3 - t_{\perp \perp}) ,
\nonumber \\
g_3 &=& 4 e_1 e_2 e_3 = 4\tau (t_{\perp \perp} -2\tau^2/9)/3 .
\end{eqnarray}
Moreover, $\wp(z;g_2,g_3)$ has douple poles at $z = 0$ and $z = L$,
because $m(z)$ has simple poles at these positions, so that the film
thickness $L$ is one of the periods of $\wp(z;g_2,g_3)$. So far our
statements are valid for both the $(+,+)$ and the $(+,-)$ boundary
condition. In order to derive the specific functional forms of the
profiles we now consider each boundary condition separately.

Turning to the $(+,+)$ boundary condition first, we note that $m'(L/2)
= 0$ whereby $t_{\perp \perp} = -\tau m^2(L/2) - m^4(L/2)$ and
\Eq{e1e2e3} simplifies to
\begin{eqnarray}
\label{e123pp}
e_1 &=& \wp(\omega_1;g_2,g_3) = P_{+,+}(L/2) = m_{+,+}^2(L/2) +
\tau/3 \nonumber \\
e_2 &=& \wp(\omega_1+\omega_2,;g_2,g_3) = \tau/3 \\
e_3 &=& \wp(\omega_2;g_2,g_3) = -P_{+,+}(L/2) - \tau/3 =
-m_{+,+}^2(L/2) - 2\tau/3 .\nonumber
\end{eqnarray}
The quantities $\omega_1$ and $\omega_2$ are the basic semiperiods
of $\wp(z;g_2,g_3)$. From \Eqs{e1e2e3}{e123pp} we conclude that
$e_1 > 0$ for all values of $\tau$ and therefore $P_{+,+}(z) =
\wp(z;g_2,g_3) > 0$ for all $0 < z < L$. Therefore, the first basic
semiperiod $\omega_1$ of the Weierstrass function can be chosen as
$\omega_1 = L/2$. It is then convenient to choose the second basic
semiperiod $\omega_2$ to be purely imaginary. We can now define the
moduli $k$ and $k'$ of the corresponding Jacobian elliptic functions
by \cite{GR80}
\begin{equation}
\label{kkp}
k^2 = {e_2 - e_3 \over e_1 - e_3} = {m^2(L/2) + \tau \over 2m^2(L/2) +
\tau} \ , \ k'^2 = 1 - k^2 .
\end{equation}
According to \Eq{kkp} bulk criticality $(\tau = 0)$ corresponds to $k^2
= k'^2 = 1/2$. The two basic semiperods are then given by the
complete elliptic integrals of the first kind $K \equiv K(k)$ and $K'
\equiv K(k')$ according to \cite{GR80}
\begin{equation}
\label{om1om2}
\omega_1 = {L \over 2} = {K \over \sqrt{e_1 - e_2}} = {K \over
\sqrt{2m^2(L/2) + \tau}}\ ,\ \omega_2 = i {K' \over K} \omega_1 .
\end{equation}
Combining \Eqs{kkp}{om1om2} we find the useful parameterization
\begin{equation}
\label{tautpp}
\tau L^2 = (2K)^2 (2k^2 - 1)\ ,\
t_{\perp \perp} = -(2K/L)^4 k^2 (1 - k^2)
\end{equation}
of the Casimir force $\langle T_{\perp \perp} \rangle_0 = (6/g)
t_{\perp \perp}$ as a function of the film thickness $L$ and the
scaling argument $\tau L^{1/\nu} = \tau L^2$ within the mean field
approximation. Finally, the order parameter function $m_{+,+}(z)$ can
be written in the form \cite{GR80}
\begin{equation}
\label{mpp}
m_{+,+}(z) = {2K \over L} {\mbox{dn}(\zeta;k) \over \mbox{sn}(\zeta;k)}\ ,
\ \zeta = {2K \over L}z ,
\end{equation}
where dn$(\zeta;k)$ and sn$(\zeta;k)$ are the Jacobian delta amplitude and
sine amplitude functions, respectively. A slight disadvantage of
\Eqs{tautpp}{mpp} is that in order to parameterize values $\tau L^2 <
-\pi^2$ one has to switch to negative values of $k^2$, i.e., to purely
imaginary moduli $k$ in the Jacobian elliptic functions dn and sn.
An alternative parameterization can be found easily by interchanging
$e_2$ and $e_3$ in \Eq{e1e2e3}. From the corresponding modification of
\Eqs{kkp}{om1om2} we find the new parameterization $(k^2 \geq 0)$
\begin{equation}
\label{tautpp1}
\tau L^2 = - (2K)^2 (k^2 + 1)\ , \ t_{\perp \perp} = (2K/L)^4 k^2
\end{equation}
for $\tau L^2 \leq -\pi^2$ and the corresponding order parameter
function reads
\begin{equation}
\label{mpp1}
m_{+,+}(z) = {2K \over L} {1 \over \mbox{sn}(\zeta;k)}\ ,
\ \zeta = {2K \over L}z .
\end{equation}
From the symmetry of the order parameter profile for $(+,+)$ boundary
conditions it is obvious that within the mean field approximation the
case of $(SB,+)$ boundary conditions can be obtained from
\Eqs{tautpp}{mpp} and their counterparts \Eqs{tautpp1}{mpp1} by the
simple transformation $L \to 2L$. The corresponding order parameter
profile is then given by $m_{+,+}(z+L)$ evaluated in the interval $0
\leq z \leq L$.

We now turn to the case of $(+,-)$ boundary conditions by noting that
in this case $m(L/2) = 0$, because $m(z)$ is antisymmetric around
$z = L/2$. Therefore, we now have $t_{\perp \perp} = m'^2(L/2)$ and
instead of \Eq{e123pp} we find
\begin{eqnarray}
\label{e123pm}
e_1 &=& \wp(\omega_1;g_2,g_3) = -\tau/6 -i \sqrt{m_{+,-}'^2(L/2)
-\tau^2/4} \nonumber \\
e_2 &=& \wp(\omega_1+\omega_2,;g_2,g_3) = P_{+,-}(L/2) = \tau/3 \\
e_3 &=& \wp(\omega_2;g_2,g_3) = -\tau/6 +i \sqrt{m_{+,-}'^2(L/2)
-\tau^2/4} \nonumber
\end{eqnarray}
indicating that this time the two basic semiperiods are complex
conjugates with $\omega_1 + \omega_2 = 2\Re\ \omega_1 = L/2$. In this
case it is convenient to define the moduli $k$ and $k'$ as \cite{GR80}
\begin{equation}
\label{kkppm}
k^2 = 1/2 - \tau/(4|m'_{+,-}(L/2)|)\ ,\ k'^2 = 1 - k^2 .
\end{equation}
The basic semiperiods can then be obtained from \cite{GR80}
\begin{equation}
\label{om1om2pm}
\omega_1 + \omega_2 = {L \over 2} = {K \over \sqrt{|m'_{+,-}(L/2)|}}\ ,
\ \omega_2 - \omega_1 = i{K' \over K} (\omega_1 + \omega_2) .
\end{equation}
Combining \Eqs{kkppm}{om1om2pm} as above we find the useful
parameterization
\begin{equation}
\label{tautpm}
\tau L^2 = -2 (2K)^2 (2k^2 - 1)\ , \ t_{\perp \perp} = (2K/L)^4
\end{equation}
of the scaling argument $\tau L^2$ and the Casimir force $\langle
T_{\perp \perp} \rangle_0$ for $(+,-)$ boundary conditions. The
corresponding order parameter function $m_{+,-}(z)$ can be written
in the form \cite{GR80}
\begin{equation}
\label{mpm}
m_{+,-}(z) = {2K \over L} {\mbox{cn}(\zeta;k) \over \mbox{sn}(\zeta;k)
\mbox{dn}(\zeta;k)}\ , \ \zeta = {2K \over L} z ,
\end{equation}
where in addition to \Eq{mpp} the Jacobian cosine amplitude cn$(\zeta;k)$
occurs. The parameterizations given by \Eqs{tautpm}{mpm} have the
disadvantage that values $\tau L^2 > 2\pi^2$ of the scaling variable
correspond to purely imaginary values of the modulus $k$. However, in
analogy with the $(+,+)$ boundary conditions the alternative
parameterization
\begin{equation}
\label{tautpm1}
\tau L^2 = 2 (2K)^2 (k^2 + 1)\ ,
\ t_{\perp \perp} = (2K/L)^4 (1 - k^2)^2
\end{equation}
can be found, where $\tau L^2 \geq 2\pi^2$ corresponds to $k^2 \geq 0$
and the corresponding expression for the profile $m_{+,-}(z)$ reads
\begin{equation}
\label{mpm1}
m_{+,-}(z) = {2K \over L} {\mbox{cn}(\zeta;k) \mbox{dn}(\zeta;k) \over
\mbox{sn}(\zeta;k)}\ , \ \zeta = {2K \over L} z.
\end{equation}
For $(O,+)$ boundary conditions the Casimir force and the profile can
be extracted from \Eqs{tautpm}{mpm} or \Eqs{tautpm1}{mpm1} by the same
simple transformation $L \to 2L$ as described above for $(SB,+)$
boundary conditions.

We close this section with the remark that the order parameter
profiles determined here can be written in the scaling form $m(z) =
L^{-\beta/\nu} h(x;y)$, where $x = z/L$ and $y = \tau L^{1/\nu}$ are
the scaling arguments and $\beta = \nu = 1/2$ within mean field
theory. The $y$ dependence of the profiles is determined by the
above parameterizations $y = y(k)$ in terms of the modulus $k$ of the
Jacobian elliptic functions. The scaling functions $h_{+,+}$ and
$h_{+,-}$ can be easily read off from \Eqs{mpp}{mpp1} and
\Eqs{mpm}{mpm1}, respectively.
\setlength{\columnwidth}{14.0cm}
\begin{figure}[t]
\centerline{\epsfbox{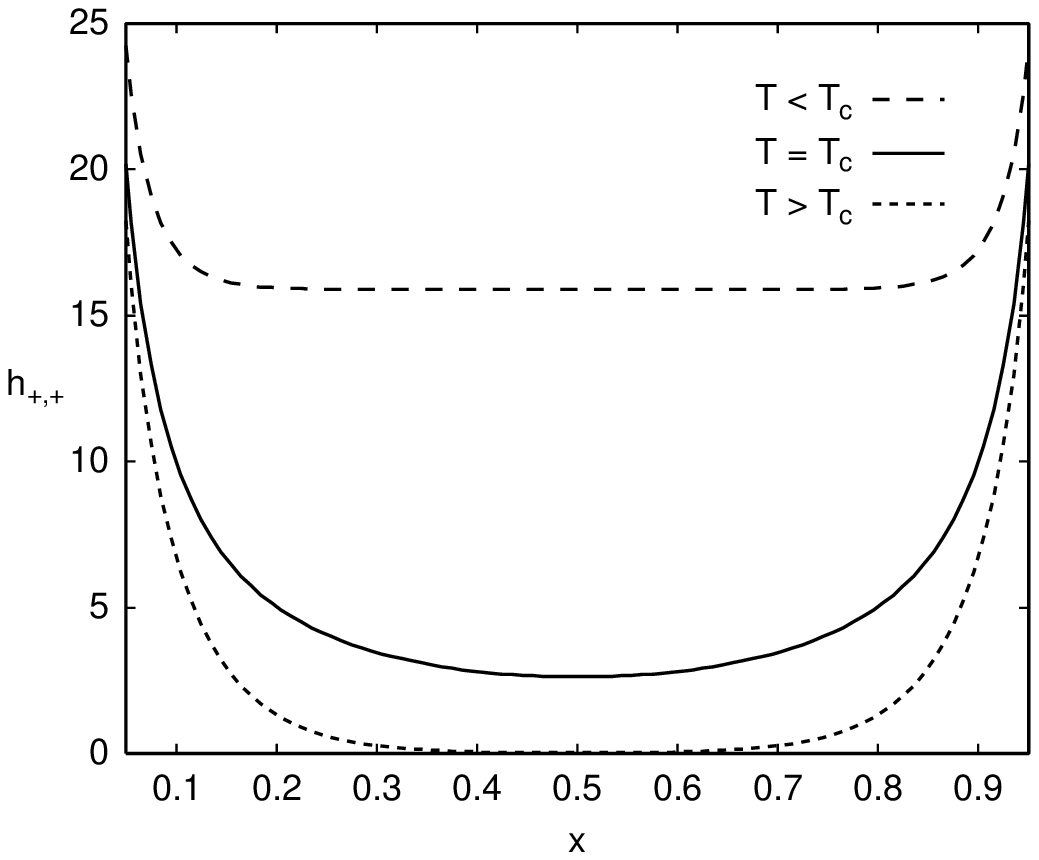}}
\centerline{\caption{\protect\small{
Scaling function $h_{+,+}(x;y)$ for $T < T_c$ $(y < 0)$ (long dashed
line), $T = T_c$ $(y = 0)$ (solid line), and $T > T_c$ $(y > 0)$
(short dashed line) according to \protect\Eq{hpp} as a function of
$x$. $T_c$ denotes the bulk critical temperature. For $y \neq 0$ the
thick-film limit $(|y| \gg 1)$ is shown (see main text).
\label{hppplot}}}}
\end{figure}
\begin{figure}[t]
\centerline{\epsfbox{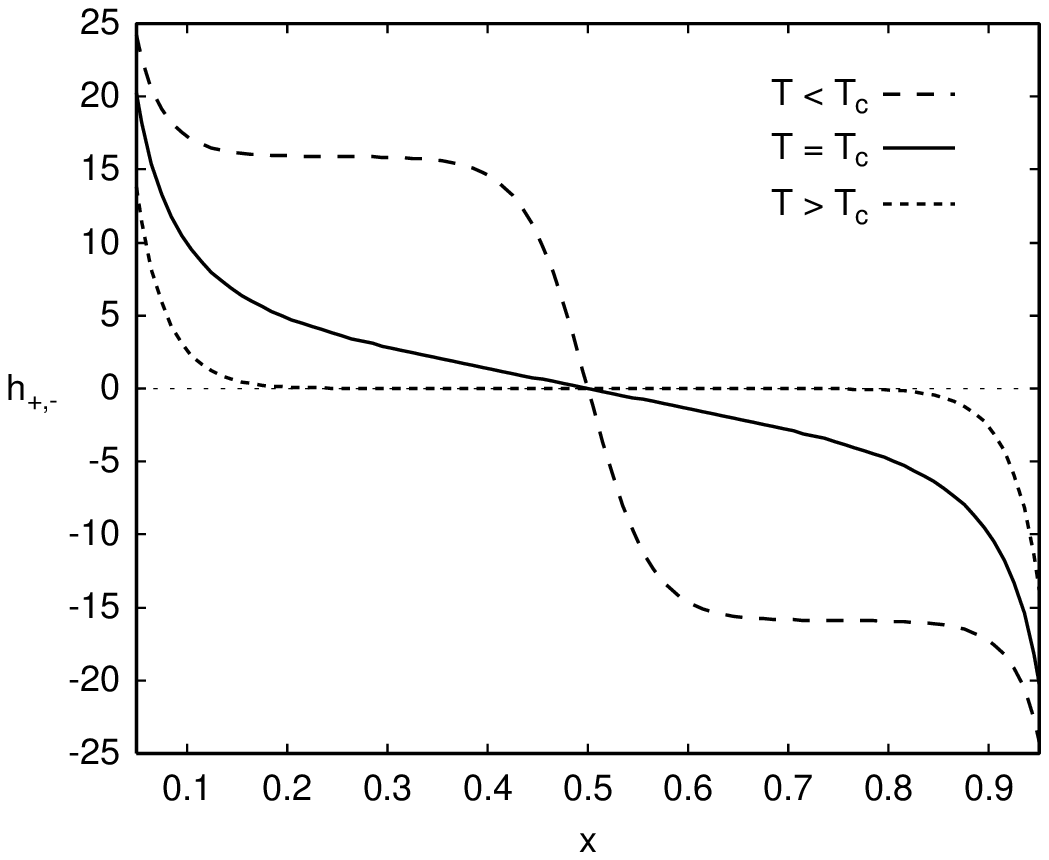}}
\centerline{\caption{\protect\small{
Scaling function $h_{+,-}(x;y)$ for $T < T_c$ $(y < 0)$ (long dashed
line), $T = T_c$ $(y = 0)$ (solid line), and $T > T_c$ $(y > 0)$
(short dashed line) according to \protect\Eq{hpm} as a function of
$x$. $T_c$ denotes the bulk critical temperature. For $y \neq 0$ the
thick-film limit $(|y| \gg 1)$ is shown (see main text).
\label{hpmplot}}}}
\end{figure}
\setlength{\columnwidth}{\oldcolwidth}
One obtains
\begin{eqnarray}
\label{hpp}
h_{+,+}(x;y) = 2K {\mbox{dn}(2Kx;k) \over \mbox{sn}(2Kx;k)}
&\quad , \quad& y = (2K)^2 (2k^2 - 1), \nonumber \\ \\
h_{+,+}(x;y) = 2K {1 \over \mbox{sn}(2Kx;k)}
&\quad , \quad& y = -(2K)^2 (k^2 + 1), \nonumber
\end{eqnarray}
and
\begin{eqnarray}
\label{hpm}
h_{+,-}(x;y) = 2K {\mbox{cn}(2Kx;k) \over
\mbox{sn}(2Kx;k) \mbox{dn}(2Kx;k)}
&\quad , \quad& y = -2 (2K)^2 (2k^2 - 1), \nonumber \\ \\
h_{+,-}(x;y) = 2K {\mbox{cn}(2Kx;k) \mbox{dn}(2Kx;k)
\over \mbox{sn}(2Kx;k)}
&\quad , \quad& y = 2 (2K)^2 (k^2 + 1). \nonumber
\end{eqnarray}
The functional forms of $h_{+,+}$ and $h_{+,-}$ below, at, and above
bulk criticality are displayed in Figs. \ref{hppplot} and
\ref{hpmplot}, respectively. Bulk criticality means $y = 0$, i.e.,
$k^2 = 1/2$ and off bulk criticality the thick film limit $|y|
\gg 1$ $(k \to 1)$ is shown. In terms of the bulk correlation length
$\xi$ the limit $|y| \gg 1$ in Figs.\ref{hppplot} and \ref{hpmplot}
is represented as $L/\xi > 15$.

\section{Eigenmode spectra}
The Gaussian Hamiltonian given by \Eq{HGauss} can be conveniently
diagonalized by solving the eigenvalue problem
\begin{equation}
\label{SEq}
-\nabla^2 \Psi({\bf x},z) + l(l + 1) m^2(z) \Psi({\bf x},z) = E
\Psi({\bf x},z) ,
\end{equation}
where $l = 1$ for the transverse spectrum and $l = 2$ for the
longitudinal spectrum and $0 \leq z \leq L$. The film geometry is
homogeneous and isotropic with respect to ${\bf x}$ so that we can
write $\Psi({\bf x},z)$ in the product form
\begin{equation}
\label{Psiprod}
\Psi({\bf x},z) = (2\pi)^{-(d-1)/2}e^{i{\bf p} \cdot {\bf x}}
\psi_n^{(l)}(z) ,
\end{equation}
where ${\bf p}$ is the longitudinal momentum and $\psi_n^{(l)}(z)$
solves the eigenvalue equation
\begin{equation}
\label{psieq}
-{d^2 \over dz^2}\psi_n^{(l)}(z) + l(l + 1) m^2(z) \psi_n^{(l)}(z) =
\epsilon_n^{(l)} \psi_n^{(l)}(z)
\end{equation}
so that the eigenvalue $E$ in \Eq{SEq} takes the form $E = {\bf p}^2 +
\epsilon_n^{(l)}$ for $l = 1$ and $l = 2$, respectively. As shown in
\Eqs{P}{e1e2e3} $m^2(z)$ is given by the Weierstrass elliptic function
$\wp(z) \equiv \wp(z;g_2,g_3)$, where $g_3 = 0$ for the case
$\tau = 0$ considered here [see \Eq{g2g3}]. Therefore \Eq{psieq} is
identical to the well known Lam{\'e} differential equation \cite{K71}
written in the form of an eigenvalue problem. The solutions of
\Eq{psieq} are known for $l = 1$ and $l = 2$ and can be used to
construct the eigenfunctions $\psi_n^{(l)}(z)$. Note that due to
$m^2(z) = \wp(z) \sim 1/z^2$ for $z \to 0$ one has $\psi_n^{(l)}(z)
\sim z^{l+1}$ for $z \to 0$ by inspection of \Eq{psieq}. Furthermore,
$(+,+)$ and $(+,-)$ boundary conditions can be treated on the same
footing by noting that according to \Eqs{om1om2}{om1om2pm} one has
\begin{eqnarray}
\label{om1om2ppm}
\omega_1^{(++)} = L/2 &\ ,\ & \omega_2^{(++)} = i L/2 , \nonumber \\
\omega_1^{(+-)} = (1-i) L/4 &\ ,\ & \omega_2^{(+-)} = (1+i) L/4
\end{eqnarray}
for the basic semiperiods of the Weierstrass function. The spectra for
the cases $(SB,+)$ and $(O,+)$ can be constructed from the spectra for
$(+,+)$ and $(+,-)$ boundary conditions, respectively.

First we turn to the transverse spectrum . According to
Ref. \cite{K71} the eigenfunctions $\psi_n^{(1)}(z)$ up to a
normalization constant can be written in the form
\begin{equation}
\label{psi1}
\psi_n^{(1)}(z) = \left[ \sigma(z+\alpha_n) e^{-z \zeta(\alpha_n)} +
\sigma(z-\alpha_n) e^{z \zeta(\alpha_n)} \right] / \sigma(z) ,
\end{equation}
where
\begin{equation}
\label{eps1}
\epsilon_n^{(1)} = - \wp(\alpha_n)
\end{equation}
yields the eigenvalues and $\zeta(z)$ and $\sigma(z)$ are the
Weierstrass $\zeta$ and $\sigma$ functions, respectively \cite{GR80}.
The spectral parameter $\alpha_n$ can be obtained from the requirement
$\psi_n^{(1)}(z) = \pm \psi_n^{(1)}(z+L)$, i.e., the eigenfunctions
are either even or odd functions when continued analytically to the
interval $[-L,L]$. From \Eq{om1om2ppm} one has $L = 2\omega_1^{(++)}
= 2(\omega_1^{(+-)} + \omega_2^{(+-)})$ and using the shift properties of
$\sigma(z)$ \cite{GR80} the above shift operation can be directly
applied to \Eq{psi1}. One obtains for the eigenvalue spectrum
\begin{equation}
\label{spec1}
2\alpha_n \zeta(L/2) - L \zeta(\alpha_n) = n \pi i\ ,\
\epsilon_n^{(1)} = -\wp(\alpha_n)\ ,\ n \geq 2 ,
\end{equation}
where the lower bound on the mode index $n$ comes from the requirement
$\psi_n^{(1)}(z) \sim z^2$ for $z \to 0$ for the transverse
eigenfunctions (see above).

For the longitudinal spectrum $(l = 2)$ the eigenfunctions take the
form \cite{K71}
\begin{equation}
\label{psi2}
\psi_n^{(2)}(z) = {d \over dz} \left\{ \left[
\sigma(z+\alpha_n) e^{-z [\zeta(\alpha_n) + \beta_n]} +
\sigma(z-\alpha_n) e^{z [\zeta(\alpha_n) + \beta_n]} \right]
/ \sigma(z) \right\} ,
\end{equation}
where
\begin{equation}
\label{eps2}
\beta_n = {\wp'(\alpha_n) \over 2 \wp(\alpha_n) +
\epsilon_n^{(2)}/3} \quad , \quad \wp(\alpha_n) =
{(\epsilon_n^{(2)})^3 \over 27 g_2 - 9 (\epsilon_n^{(2)})^2}
\end{equation}
yields the eigenvalues and $\wp'(z)$ denotes the derivative of
the Weierstrass $\wp$-function with respect to $z$.
We again employ the symmetry requirement $\psi_n^{(2)}(z) = \pm
\psi_n^{(2)}(z+L)$ and the boundary behavior $\psi_n^{(2)}(z) \sim
z^3$ for $z \to 0$ to obtain
\begin{equation}
\label{spec2}
2\alpha_n \zeta(L/2) - L \left[\zeta(\alpha_n)+{\wp'(\alpha_n) \over
2 \wp(\alpha_n) + \epsilon_n^{(2)}/3} \right] = n \pi i\ ,\ 
\wp(\alpha_n) =
{(\epsilon_n^{(2)})^3 \over 27 g_2 - 9 (\epsilon_n^{(2)})^2}\ ,\
n \geq 3 .
\end{equation}
The solution of \Eqs{spec1}{spec2} for the eigenvalues
$\epsilon_n^{(i)}$, $i=1,2$ cannot be obtained in a closed analytic
form. In order to deal with the divergencies of the mode sums in
\Eqs{FGauss}{TppGauss} (see also Appendix C), we derive the asymptotic
behavior of the eigenvalues from \Eqs{spec1}{spec2} for large $n$.
From the geometry of the problem it is clear that the leading term in
an expansion of $\epsilon_n^{(i)}$ in powers of $1/n$ is given by
the spectrum $(n\pi/L)^2$ of a free particle in a one-dimensional box
of length $L$. Therefore the spectral parameter $\alpha_n$ behaves as
$1/n$ as $n$ increases, so that the desired asymptotic form of the $n$
dependence of the eigenvalues can be obtained from \Eqs{spec1}{spec2}
by expanding the Weierstrass functions $\zeta(\alpha_n)$,
$\wp(\alpha_n)$, and $\wp'(\alpha_n)$ in powers of $\alpha_n$, where
only the leading two terms are needed. Specifically, we use the
expansions \cite{GR80}
\begin{equation}
\label{expans}
\zeta(x) = 1/x - g_2/60\ x^3 + {\cal O}(x^7)\ ,\
\wp(x) = 1/x^2 + g_2/20\ x^2 + {\cal O}(x^6) ,
\end{equation}
where $g_3 = 0$ is implicitly assumed. The calculation is
straightforward so that we only briefly summarize the results for the
eigenvalues $\epsilon_n^{(i)}$. Corresponding expansions are obtained
for the spectral parameter $\alpha_n$ which will not be reproduced
here.

For $(+,+)$ boundary conditions one has
\begin{equation}
\label{pppar}
\zeta(L/2) = \zeta(\omega_1^{(++)}) = \pi/(2L)\ ,\ g_2 = -4 t_{\perp
\perp} = (2K/L)^4 ,
\end{equation}
where $K = K(1/\sqrt{2})$ [see \Eqss{g2g3}{tscal}{fpp}]. By insertion
of \Eqs{expans}{pppar} into \Eqs{spec1}{spec2} one obtains the
expansions
\begin{eqnarray}
\label{specapp}
\epsilon_n^{(1)} &=& \left({n \pi \over L}\right)^2
\left[1 - {2 \over \pi n^2} + {1 \over \pi^2 n^4}
\left({4 K^4 \over 3 \pi^2} - 1 \right) + {\cal O}(n^{-6}) \right]\ ,\
(n \geq 2) \nonumber \\ \\
\epsilon_n^{(2)} &=& \left({n \pi \over L}\right)^2
\left[1 - {6 \over \pi n^2} + {9 \over \pi^2 n^4}
\left({4 K^4 \over 3 \pi^2} - 1 \right) + {\cal O}(n^{-6})\right]\ ,\
(n \geq 3) . \nonumber
\end{eqnarray}
For $(+,-)$ boundary conditions one has correspondingly
\begin{equation}
\label{pmpar}
\zeta(L/2) = \zeta(\omega_1^{(+-)}+\omega_2^{(+-)}) = \pi/L\ ,\
g_2 = -4 t_{\perp \perp} = -4 (2K/L)^4 ,
\end{equation}
where $K$ is given as above [see \Eqss{g2g3}{tscal}{fpm}]. Insertion
of \Eqs{expans}{pmpar} into \Eqs{spec1}{spec2} yields the expansions
\begin{eqnarray}
\label{specapm}
\epsilon_n^{(1)} &=& \left({n \pi \over L}\right)^2
\left[1 - {4 \over \pi n^2} - {4 \over \pi^2 n^4}
\left({4 K^4 \over 3 \pi^2} + 1 \right) + {\cal O}(n^{-6}) \right]\ ,\
(n \geq 2) \nonumber \\ \\
\epsilon_n^{(2)} &=& \left({n \pi \over L}\right)^2
\left[1 - {12 \over \pi n^2} - {36 \over \pi^2 n^4}
\left({4 K^4 \over 3 \pi^2} + 1 \right) + {\cal O}(n^{-6}) \right]\ ,\
(n \geq 3) . \nonumber
\end{eqnarray}
The asymptotic expressions for the spectrum given by
\Eqs{specapp}{specapm} capture all divergent terms in the mode sums in
\Eqs{FGauss}{TppGauss} as will be seen in Appendix C. Furthermore,
\Eqs{specapp}{specapm} provide very good initial values for a numerical
solution of \Eqs{spec1}{spec2} by iterative schemes, e.g., the Newton
procedure.

For $(SB,+)$ boundary conditions the eigenvalue spectra can be
obtained from the case of $(+,+)$ boundary conditions by employing the
transformation $L \to 2L$ and by allowing only {\em even} indices $n$
for $\epsilon_n^{(1)}$ and only {\em odd} indices $n$ for
$\epsilon_n^{(2)}$ [see \Eq{specaSBp}]. Likewise, the eigenvalue
spectra for $(O,+)$ boundary conditions can be obtained from the case
of $(+,-)$ boundary conditions by again employing the transformation
$L \to 2L$ and by allowing only {\em odd} indices $n$ for
$\epsilon_n^{(1)}$ and only {\em even} indices $n$ for
$\epsilon_n^{(2)}$ [see \Eq{specaOp}]. The reason for this simple rule
is that for $(+,+)$ boundary conditions starting from the ground state
every second eigenfunction has vanishing slope at $z = L/2$ so that
after rescaling $L \to 2L$ the eigenfunctions for $(SB,+)$ boundary
conditions are already contained in the $(+,+)$ case. An analogous
argument relates the spectra for $(+,-)$ and $(O,+)$ boundary
conditions starting from the first excited state for the $(+,-)$ case.

\section{Regularized mode sums}
The mode sums appearing in \Eqs{FGauss}{TppGauss} are divergent for
any spatial dimension $d$ of interest. Within the dimensional
regularization scheme used throughout this investigation $d$ is used
as a free parameter in order to find an analytic continuation of the
mode sums as a function of $d$, where $d = 4 - \varepsilon$ is this
case. On the other hand the mode sums in \Eqs{FGauss}{TppGauss} also
constitute the zeta functions of the eigenvalue spectrum with a $d$
dependent argument \cite{E94}. The zeta function regularization of
mode sums, which is a widely used technique to treat divergent series
like those in \Eqs{FGauss}{TppGauss} \cite{E94}, is therefore
equivalent to the dimensional regularization scheme.

The major obstacle towards an analytical treatment of the
aforementioned mode sums has been removed in Appendix B by the
derivation of the asymptotic behavior of the eigenvalue spectrum for
large mode numbers given by \Eqs{specapp}{specapm}. Using these
results one has for $i=1,2$
\begin{equation}
\label{restsum}
\sum_{n=n_0}^\infty \left\{(\epsilon_n^{(i)})^{(d-1)/2} - \left[{n \pi
\over L} \right]^{d-1} \left[1 - {2A \over n^2} + {B \over n^4}
\right]^{(d-1)/2} \right\} \sim \sum_{n=n_0}^\infty n^{d-7}
\end{equation}
which is convergent for any $d$ of physical interest and can thus be
determined numerically from the solutions of \Eqs{spec1}{spec2} for
the transverse and the longitudinal mode sum, respectively. The
problem of regularizing the mode sums has therefore reduced to the
regularization of the corresponding sums over the large $n$ expansions
given by \Eqs{specapp}{specapm}, i.e., one has to consider the series 
\begin{equation}
\label{aseries}
\sum_{n=n_0}^\infty n^{d-1}
\left[1 - {2A \over n^2} + {B \over n^4} \right]^{(d-1)/2}
\end{equation}
for $d = 4 - \varepsilon$. 
If the lower summation bound $n_0$ in \Eq{aseries} is chosen
sufficiently large, one can safely expand the term under the sum in
powers of $1/n^2$ which leads to an expansion of the series given by
\Eq{aseries} in terms of Hurwitz functions $\zeta(x,n_0)$. One finds
for $d = 4 - \varepsilon$
\begin{eqnarray}
\label{regser}
\sum_{n=n_0}^\infty
\left[n^2 - 2A + {B \over n^2} \right]^{(3-\varepsilon)/2}
&=& \zeta(-3,n_0) - 3A \zeta(-1,n_0) \nonumber \\ \nonumber \\
&+& {3-\varepsilon \over 2} \left[(1-\varepsilon) A^2 + B\right]
\zeta(1+\varepsilon,n_0) \nonumber \\ \nonumber \\
&+& {A \over 2} (A^2 - 3B) \zeta(3,n_0) + {3 \over 8} (A^2 - B)^2
\zeta(5,n_0) \nonumber \\ \nonumber \\
&+& {3 \over 8} A (A^2 - B)^2 \zeta(7,n_0) + {\cal O}(\varepsilon) +
{\cal O}(1/n_0^8) ,
\end{eqnarray}
where the $\varepsilon$-expansion has already been carried out up to
terms ${\cal O}(\varepsilon)$. The expansion shown in \Eq{regser}
converges quite fast already for $3 \leq n_0 \leq 5$. The
$1/\varepsilon$ pole indicating the UV singularity can be extracted
from \Eq{regser} using the expansion
\begin{equation}
\label{zeta1}
\zeta(1+\varepsilon,n_0) = 1/\varepsilon + \gamma - \sum_{k=1}^{n_0-1}
1/k + {\cal O}(\varepsilon) ,
\end{equation}
where $\gamma \simeq 0.577216$ is the Euler constant and $n_0$ is a
positive integer. With the coefficients $A$ and $B$ taken from
\Eqs{specapp}{specapm} the expressions given by
\Eqss{restsum}{regser}{zeta1} can be combined to the following
regularized and $\varepsilon$-expanded expressions for the mode sums.

For $(+,+)$ boundary conditions one finds with $K \equiv K(1/\sqrt{2})$
\begin{eqnarray}
\label{sumpp}
\sum_{n=2}^\infty \left( \epsilon_n^{(1)} \right)^{(d-1)/2}
&=& \left[\pi \over L\right]^{d-1} {2 K^4 \over \pi^4 \varepsilon}
\left[1 - \varepsilon \left({3 \pi^2 \over 4 K^4} + {1 \over 3} -
\gamma + 0.7494 \right)\right] + {\cal O}(\varepsilon), \nonumber \\ \\
\sum_{n=3}^\infty \left( \epsilon_n^{(2)} \right)^{(d-1)/2}
&=& \left[\pi \over L\right]^{d-1} {18 K^4 \over \pi^4 \varepsilon}
\left[1 - \varepsilon \left({3 \pi^2 \over 4 K^4} + {1 \over 3} -
\gamma + 1.5589 \right)\right] + {\cal O}(\varepsilon). \nonumber
\end{eqnarray}
For $(+,-)$ boundary conditions the corresponding result reads
\begin{eqnarray}
\label{sumpm}
\sum_{n=2}^\infty \left( \epsilon_n^{(1)} \right)^{(d-1)/2}
&=& -\left[\pi \over L\right]^{d-1} {8 K^4 \over \pi^4 \varepsilon}
\left[1 + \varepsilon \left({3 \pi^2 \over 4 K^4} - {1 \over 3} +
\gamma - 1.7198 \right)\right] + {\cal O}(\varepsilon), \nonumber \\ \\
\sum_{n=3}^\infty \left( \epsilon_n^{(2)} \right)^{(d-1)/2}
&=& -\left[\pi \over L\right]^{d-1} {72 K^4 \over \pi^4 \varepsilon}
\left[1 + \varepsilon \left({3 \pi^2 \over 4 K^4} - {1 \over 3} +
\gamma - 2.4086 \right)\right] + {\cal O}(\varepsilon). \nonumber
\end{eqnarray}
For $(SB,+)$ boundary conditions we apply the simple transformation
described in the last paragraph of Appendix B to the eigenvalue
spectrum for $(+,+)$ boundary conditions. From \Eq{specapp} we find
the expansions
\begin{eqnarray}
\label{specaSBp}
\epsilon_n^{(1)} &=& \left({n \pi \over L}\right)^2
\left[1 - {1 \over 2\pi n^2} + {1 \over 16 \pi^2 n^4}
\left({4 K^4 \over 3 \pi^2} - 1 \right) + {\cal O}(n^{-6}) \right]\ ,\
(n \geq 1) \nonumber \\ \\
\epsilon_n^{(2)} &=& \left({(2n+1) \pi \over 2L}\right)^2
\left[1 - {6 \over \pi (2n+1)^2} + {9 \over \pi^2 (2n+1)^4}
\left({4 K^4 \over 3 \pi^2} - 1 \right) + {\cal O}(n^{-6}) \right]\ ,\
(n \geq 1) , \nonumber
\end{eqnarray}
for the transverse and the longitudinal spectrum, respectively. Due to
the appearance of half-integer arguments in the Hurwitz functions for
the transverse mode sum in this case one needs the expansion
\begin{equation}
\label{zeta12}
\zeta(1+\varepsilon,3/2) = 1/\varepsilon + \gamma + 2\ln 2 - 2 +
{\cal O}(\varepsilon)
\end{equation}
instead of \Eq{zeta1}. Furthermore, the r.h.s. of \Eq{regser} with
$n_0$ replaced by $3/2$ is needed in order to derive the regularized
longitudinal mode sum. The transverse mode sum, however, can be
evaluated directly using \Eqs{regser}{zeta1}. One therefore finds for
$(SB,+)$ boundary conditions
\begin{eqnarray}
\label{sumSBp}
\sum_{n=1}^\infty \left( \epsilon_n^{(1)} \right)^{(d-1)/2}
&=& \left[\pi \over L\right]^{d-1} {K^4 \over 8 \pi^4 \varepsilon}
\left[1 - \varepsilon \left({3 \pi^2 \over 4 K^4} + {1 \over 3} -
\gamma - 1.9712 \right)\right] + {\cal O}(\varepsilon), \nonumber \\ \\
\sum_{n=1}^\infty \left( \epsilon_n^{(2)} \right)^{(d-1)/2}
&=& \left[\pi \over L\right]^{d-1} {9 K^4 \over 8 \pi^4 \varepsilon}
\left[1 - \varepsilon \left({3 \pi^2 \over 4 K^4} + {1 \over 3} -
\gamma - 0.8405 \right)\right] + {\cal O}(\varepsilon). \nonumber
\end{eqnarray}
For $(O,+)$ boundary conditions we apply the same transformation
to the eigenvalue spectrum for $(+,-)$ boundary conditions. From
\Eq{specapm} we find the expansions
\begin{eqnarray}
\label{specaOp}
\epsilon_n^{(1)} &=& \left({(2n+1)\pi \over 2L}\right)^2
\left[1 - {4 \over \pi (2n+1)^2} - {4 \over \pi^2 (2n+1)^4}
\left({4 K^4 \over 3 \pi^2} + 1 \right) + {\cal O}(n^{-6}) \right]\ ,\
(n \geq 1) \nonumber \\ \\
\epsilon_n^{(2)} &=& \left({n \pi \over L}\right)^2
\left[1 - {3 \over \pi n^2} - {9 \over 4 \pi^2 n^4}
\left({4 K^4 \over 3 \pi^2} + 1 \right) + {\cal O}(n^{-6}) \right]\ ,\
(n \geq 2) , \nonumber
\end{eqnarray}
Using \Eq{regser} with $n_0$ replaced by $3/2$ in order to evaluate the
transverse mode sum one therefore finds for $(O,+)$ boundary conditions
\begin{eqnarray}
\label{sumOp}
\sum_{n=1}^\infty \left( \epsilon_n^{(1)} \right)^{(d-1)/2}
&=& -\left[\pi \over L\right]^{d-1} {K^4 \over 2 \pi^4 \varepsilon}
\left[1 + \varepsilon \left({3 \pi^2 \over 4 K^4} - {1 \over 3} +
\gamma - 1.8975 \right)\right] + {\cal O}(\varepsilon), \nonumber \\ \\
\sum_{n=2}^\infty \left( \epsilon_n^{(2)} \right)^{(d-1)/2}
&=& -\left[\pi \over L\right]^{d-1} {9 K^4 \over 2 \pi^4 \varepsilon}
\left[1 + \varepsilon \left({3 \pi^2 \over 4 K^4} - {1 \over 3} +
\gamma - 1.9276 \right)\right] + {\cal O}(\varepsilon). \nonumber
\end{eqnarray}
In order to facilitate the $\varepsilon$-expansion of \Eq{TppGauss} we
finally note that
\begin{equation}
\label{G3d2}
\Gamma\left[(3-d)/2\right] = -2\sqrt{\pi} \left[1 + \varepsilon (1 -
\ln 2 - \gamma/2) + {\cal O}(\varepsilon^2) \right]
\end{equation}
and
\begin{equation}
\label{Gd2}
\Gamma(d/2) = 1 - \varepsilon (1 - \gamma)/2 + {\cal O}(\varepsilon^2)
\end{equation}
for $d = 4 - \varepsilon$ (see also Ref. \cite{MKSD92} for similar
relations).


\begin{thebibliography}{99}

\bibitem{KB83}
 K. Binder in {\em Phase Transitions and Critical Phenomena}, edited
 by C. Domb and J.L. Lebowitz (Academic, London, 1983), Vol.8, p. 2.
\bibitem{HWD86}
 H.W. Diehl in {\em Phase Transitions and Critical Phenomena}, edited
 by C. Domb and J.L. Lebowitz (Academic, London, 1986), Vol.10, p. 76.
\bibitem{SD88}
 S. Dietrich in {\em Phase Transitions and Critical Phenomena}, edited
 by C. Domb and J.L. Lebowitz (Academic, London, 1988), Vol.12, p. 1.
\bibitem{MEFHAY80}
 M.E. Fisher and H. Au-Yang, Physica A {\bf 101A}, 255 (1980).
\bibitem{HWDAC91}
 H.W. Diehl and A. Ciach, Phys. Rev. B {\bf 44}, 6642 (1991).
\bibitem{HWDMS93}
 H.W. Diehl and M. Smock, Phys. Rev. B {\bf 47}, 5841 (1993).
\bibitem{MEF70}
 M.E. Fisher, in {\em Proceedings of the 1970 Enrico Fermi School of
 Physics, Varenna, Italy}, Course No. LI, edited by M.S. Green
 (Academic, New York, 1971), p. 1; M. E. Fisher and M.N. Barber, Phys.
 Rev. Lett. {\bf 28}, 1516 (1972).
\bibitem{MEFHN81}
 M.E. Fisher and H. Nakanishi, J. Chem. Phys. {\bf 75}, 5857 (1981).
\bibitem{MNB83}
 M.N. Barber, in {\em Phase Transitions and Critical Phenomena},
 edited by C. Domb and J.L. Lebowitz (Academic, New York, 1983), Vol.
 8, p. 145; V. Privman, in {\em Finite Size Scaling and Numerical
 Simulation of Statistical Systems}, edited by V. Privman (World
 Scientific, Singapore, 1990).
\bibitem{MPNJOI85}
 M.P. Nightingale and J.O. Indekeu, Phys. Rev. Lett. {\bf 54}, 1824
 (1985) and Phys. Rev. Lett. {\bf 55}, 1700 (1985); R. Lipowski and U.
 Seifert, Phys. Rev. B {\bf 31}, 4701 (1985) and Phys. Rev. Lett. {\bf
 55}, 1699 (1985).
\bibitem{HNMEF83}
 H. Nakanishi and M.E. Fisher, J. Chem. Phys. {\bf 79}, 3279 (1983).
\bibitem{KBDPL91}
 K. Binder and D.P. Landau, Physica A {\bf 177}, 483 (1991).
\bibitem{REUBM86}
 R. Evans, U. Marini Bettolo Marconi, and P. Tarazona, J. Chem. Phys.
 {\bf 84}, 2376 (1986).
\bibitem{AOPRE92}
 A.O. Parry and R. Evans, J. Phys. A {\bf 25}, 275 (1992).
\bibitem{KBDPL92}
 K. Binder and D.P. Landau, J. Chem. Phys. {\bf 96}, 1444 (1992).
\bibitem{MRSALOJOI91}
 M.R. Swift, A.L. Owczarek, and J.O. Indekeu, Europhys. Lett. {\bf
 14}, 475 (1991).
\bibitem{AOPRE92a}
 A.O. Parry and R. Evans, Physica A {\bf 181}, 250 (1992).
\bibitem{KBDPLAMF95}
 K. Binder, D.P. Landau, and A.M. Ferrenberg, Phys. Rev. Lett. {\bf
 74}, 298 (1995).
\bibitem{AOP92}
 A.O. Parry, J. Phys. A {\bf 25}, 257 (1992).
\bibitem{KBREDPLAMF96}
 K. Binder, R. Evans, D.P. Landau, and A.M. Ferrenberg, Phys. Rev. E
 {\bf 53}, 5023 (1996).
\bibitem{MK94}
 M. Krech, {\em The Casimir Effect in Critical Systems} (World
 Scientific, Singapore, 1994) and references therein.
\bibitem{LS96}
 L. Spruch, Science {\bf 272}, 1452 (1996) and references therein.
\bibitem{CCS94}
 S.L. Carnie, D.Y.C. Chan, and J. Stankovich, J. Coll. Int. Sci. {\bf 169},
 116 (1994).
\bibitem{GHNIB94}
 G.H. Nyland and I. Brevik, Physica A {\bf 202}, 81 (1994); I. Brevik
 and G.H. Nyland, Ann. Phys. {\bf 230}, 321 (1994).
\bibitem{SLAR96}
 S. Leseduarte and A. Romeo, Ann. Phys. {\bf 250}, 448 (1996).
\bibitem{NSC92}
 M.Y. Novikov, A.S. Sorin, and V.Y. Chernyak, Theor. Math. Phys. {\bf 91},
 658 (1992) and {\bf 92}, 773 (1993).
\bibitem{BKM95}
 M. Borday, G.L. Klimchitskaya, and V.M. Mostepanenko, Phys. Lett. A
 {\bf 200}, 95 (1995).
\bibitem{MKSD92}
 M. Krech and S. Dietrich, Phys. Rev. A {\bf 46}, 1886 (1992).
\bibitem{JLC86}
 J.L. Cardy, Nucl. Phys. {bf B275}, 200 (1986).
\bibitem{REJS94}
 R. Evans and J. Stecki, Phys. Rev. B {\bf 49}, 8842 (1994).
\bibitem{DD96}
 D. Danchev, Phys. Rev. E {\bf 53}, 2104 (1996).
\bibitem{INW86}
 J.O. Indekeu, M.P. Nightingale, and W.V. Wang, Phys. Rev. B {\bf 34},
 330 (1986).
\bibitem{MKDPL96}
 M. Krech and D.P. Landau, Phys. Rev. E {\bf 53}, 4414 (1996).
\bibitem{TWBEEUR95}
 T.W. Burkhardt and E. Eisenriegler, Phys. Rev. Lett. {\bf 74}, 3189
 (1995); E. Eisenriegler and U. Ritschel, Phys. Rev. B {\bf 51},
 13717 (1995).
\bibitem{DOCCRS94}
 D. O'Connor and C.R. Stephens, Phys. Rev. Lett. {\bf 72}, 506 (1994);
 F. Freire, D. O'Connor, and C.R. Stephens, J. Stat. Phys. {\bf 74},
 219 (1994).
\bibitem{Dohm1}
 A. Esser, V. Dohm, and X.S. Chen, Physica A {\bf 222}, 355 (1995);
 A. Esser, V. Dohm, M. Hermes, and J.S. Wang, Z. Phys. B {\bf 97}, 205
 (1995); X.S. Chen, V. Dohm, and A.L. Talapov, Physica A {\bf 232}, 375
 (1996).
\bibitem{DohmN}
 X.S. Chen, V. Dohm, and A. Esser, J. Phys. I France {\bf 5}, 205 (1995);
 X.S. Chen, V. Dohm, and N. Schultka, Phys. Rev. Lett. {\bf 77}, 3641
 (1996).
\bibitem{MIKANO72}
 M.I. Kaganov and A.N. Omel'yanchuk, Sov. Phys. JETP {\bf 34}, 895 (1972);
 M.I. Kaganov, Sov. Phys. JETP {\bf 35}, 631 (1972).
\bibitem{BKJZJ93}
 E. Br{\'e}zin, F. Korutcheva, T. Jolicoeur, and J. Zinn-Justin, J. Stat.
 Phys. {\bf 70}, 583 (1993).
\bibitem{SGUR95}
 S. Gnutzmann and U. Ritschel, Z. Phys. B {\bf 96}, 391 (1995).
\bibitem{GG83}
 G. Gumbs, J. Math. Phys. {\bf 24}, 202 (1983).
\bibitem{TWBHWD94}
 T.W. Burkhardt and H.W. Diehl, Phys. Rev. B {\bf 50}, 3894 (1994).
\bibitem{CRFW95}
 C. Ruge, S. Dunkelmann, and F. Wagner, Phys. Rev. Lett. {\bf 69},
 2465 (1992); C. Ruge and F. Wagner, Phys. Rev. B {\bf 52}, 4209
 (1995).
\bibitem{EEMS94}
 E. Eisenriegler and M. Stapper, Phys. Rev. B {\bf 50}, 10009 (1994).
\bibitem{JRHGDJ85}
 J. Rudnick, H. Guo, and D. Jasnow, J. Stat. Phys. {\bf 41}, 353 (1985).
\bibitem{KNSCL91}
 H. Kleinert, J. Neu, V. Schulte-Frohlinde, K.G. Chetyrkin, and S.A.
 Larin, Phys. Lett. B {\bf 272}, 39 (1991).
\bibitem{AMFRHS89}
 A.M. Ferrenberg and R.H. Swendsen, Phys. Rev. Lett. {\bf 63}, 1195
 (1989).
\bibitem{MKSD92a}
 M. Krech and S. Dietrich, Phys. Rev. A {\bf 46}, 1922 (1992).
\bibitem{SDAL89}
 S. Dietrich and A. Latz, Phys. Rev. B {\bf 40}, 9204 (1989).
\bibitem{ESSCHA73}
 E.S. Sabisky and C.H. Anderson, Phys. Rev. A {\bf 7}, 790 (1973).
\bibitem{AMBML96}
 A. Mukhopadhyay and B.M. Law, private communication.
\bibitem{JNIPMM88}
 J.N. Israelachvili and P.M. McGuggian, Science {\bf 241}, 795 (1988)
 and references therein.
\bibitem{GR80}
 L.S. Gradsteyn and I.M. Ryzhik, {\em Table of Integrals, Series and
 Products} (Academic, New York, 1980), pp. 904.
\bibitem{K71}
 E. Kamke, {\em Differentialgleichungen, L\"osungsmethoden und
 L\"osungen} (Chelsea, New York, 1971), Vol. 1, pp. 408.
\bibitem{E94}
 E. Elizalde, J. Phys. A {\bf 27}, L299 (1994) and references
 therein.

\end{thebibliography}
\end{document}